\documentclass[aps,prb,twocolumn,showpacs,groupedaddress,citeautoscript]{revtex4-1}
\usepackage{graphicx,amsmath,amssymb}

\def\bm#1{\mbox{\boldmath $#1$}}
\makeatletter
\def\lsim{\mathrel{\mathpalette\gl@align<}}
\def\gsim{\mathrel{\mathpalette\gl@align>}}
\def\gl@align#1#2{\lower.6ex\vbox{\baselineskip\z@skip\lineskip\z@
    \ialign{$\m@th#1\hfil##\hfil$\crcr#2\crcr\sim\crcr}}}
\makeatother

\begin{document}

\title{Valley coupling in finite-length metallic single-wall carbon nanotubes}

\author{W. Izumida}
\email[]{izumida@cmpt.phys.tohoku.ac.jp}
\affiliation{Department of Physics, Tohoku University, Sendai
  980-8578, Japan}
\author{R. Okuyama}
\affiliation{Department of Physics, Tohoku University, Sendai 980-8578, Japan}
\author{R. Saito}
\affiliation{Department of Physics, Tohoku University, Sendai 980-8578, Japan}

\date{
}

\begin{abstract}
  Degeneracy of discrete energy levels of finite-length, metallic
  single-wall carbon nanotubes depends on type of nanotubes, boundary
  condition, length of nanotubes and spin-orbit interaction.
  Metal-1 nanotubes, in which two non-equivalent valleys in the
  Brillouin zone have different orbital angular momenta with respect
  to the tube axis, exhibits nearly fourfold degeneracy and small lift
  of the degeneracy by the spin-orbit interaction reflecting the
  decoupling of two valleys in the eigenfunctions.
  In metal-2 nanotubes, in which the two valleys have the same orbital
  angular momentum, vernier-scale-like spectra appear for boundaries
  of orthogonal-shaped edge or cap-termination reflecting the strong
  valley coupling and the asymmetric velocities of the Dirac states.
  Lift of the fourfold degeneracy by parity splitting overcomes the
  spin-orbit interaction in shorter nanotubes with a so-called minimal
  boundary.
  Slowly decaying evanescent modes appear in the energy gap induced by
  the curvature of nanotube surface.
  Effective one-dimensional model reveals the role of boundary on the
  valley coupling in the eigenfunctions.
\end{abstract}



\maketitle

\section{Introduction}

Metallic single-wall carbon nanotubes (m-SWNTs) are ideal
one-dimensional (1D) conductors of nanometer to micrometer length.
Due to the confinement in the finite-length, energy levels of
electrons are quantized and the eigenfunctions show standing wave
behavior.~\cite{Venema-1999-01,Lemay:2001aa}
Fourfold degeneracy of the discrete energy levels observed in the
tunneling spectroscopy measurements has been considered as an
intrinsic property of SWNTs reflecting the two non-equivalent,
degenerate valleys at the $K$ and $K'$ points in the two-dimensional
(2D) Brillouin zone (BZ) together with two spins degrees of
freedom.~\cite{Liang-2002-03,Cobden-2002-07,Jarillo-Herrero-2005-04,Moriyama-2005-05,Sapmaz-2005-04,Maki-2005-06,Cao-2005-09,Makarovski-2006-10,Moriyama-2007-04,Holm-2008-04}
Recent measurements with ultraclean SWNTs have found fine structures
of the order of sub-milli-electron-volt in tunneling conductance
spectra caused by spin-orbit
interaction,~\cite{Kuemmeth-2008-03,Jhang-2010-07,Jespersen-2011-04,Steele:2013fk}
which lifts the fourfold degeneracy by spin splitting in each
valley.~\cite{Ando-2000-06,Chico-2004-10,Huertas-Hernando-2006-10,Chico-2009-06,izumida-2009-06,Jeong-2009-08}
On the other hand, some experiments show degeneracy behaviors other
than above, such as gate-dependent oscillation of twofold and fourfold
degeneracy,~\cite{Maki-2005-06,Makarovski-2006-10,Moriyama-2007-04,Holm-2008-04}
and many of them have been owed to extrinsic effects such as
impurities.

In our previous study,~\cite{PhysRevB.85.165430} we pointed out the
asymmetric velocities in the same valley for the m-SWNTs because of
the curvature of nanotube surface.
That is, the left- and right-going waves in the same valley have
different velocities, $v_{\rm L}^{(K)} > v_{\rm R}^{(K)} ( > 0)$,
where $v_{\rm L}^{(K)}$ ($v_{\rm R}^{(K)}$) is the velocity of
left-going (right-going) wave at the $K$ valley, and we have the
relations $v_{\rm L}^{(K)} = v_{\rm R}^{(K')}$ and $v_{\rm L}^{(K')} =
v_{\rm R}^{(K)}$ because of the time-reversal symmetry.
At the same time, we claimed the formation of vernier-scale-like
energy spectrum, in which two sets of the energy levels with a
constant energy separation between the levels have a similar but
not-exactly the same separation for each set, if the strong valley
coupling occurs, in which each set of the wavefunction is formed from
a left-going wave at one valley and a right-going wave at another
valley.
As the result of the quantization of the wavenumber in the axis
direction, there are two different sets of equi-spaced discrete energy
levels, $\hbar v_{\rm L}^{(K)} \pi / L_{\rm NT}$ and $\hbar v_{\rm
  R}^{(K)} \pi / L_{\rm NT}$, like the vernier
scale,~\cite{PhysRevB.85.165430} showing two- and fourfold
oscillations as observed in the
experiments,~\cite{Maki-2005-06,Makarovski-2006-10,Moriyama-2007-04,Holm-2008-04}
where $L_{\rm NT}$ is the nanotube length.
On the other hand, for the case of valley decoupling, in which each
wavefunction is formed from a left- and right-going waves in the same
valley, the fourfold degeneracy and its lift by the spin-orbit
interaction~\cite{Ando-2000-06,Chico-2004-10,Huertas-Hernando-2006-10,Chico-2009-06,izumida-2009-06,Jeong-2009-08}
would be observed.
Thus, it is important to reveal the coupling of the two valleys as a
function SWNT chirality or the boundary shape for understanding the
degeneracy behavior.

As is known that the particle-in-a-1D-box model cannot be directly
applied to the m-SWNTs because there are two left-going waves and two
right-going waves,
in general, the ratios of 
these traveling waves in a standing wave are determined by microscopic
conditions such as the chirality and the boundary condition.
Previous calculations have shown the standing waves oscillating in the
length scale of carbon-carbon bond for the armchair nanotubes, in
which each standing wave is constructed in the condition of strong
valley coupling.~\cite{Rubio-1999-04,Yaguchi-2001-05}
On the other hand for the zigzag nanotubes, the slowly oscillating
standing waves for doubly degenerate levels can be constructed in the
condition of valley
decoupling.~\cite{doi:10.1021/jp972300h,PhysRevB.65.165431}
The SWNTs have been classified in terms of the boundary condition.
Using generalized parameters, McCann and Fal'ko classified the
boundary conditions for the Dirac electrons in the
m-SWNTs.~\cite{McCann-2004-04}
By employing microscopic analysis on the boundary modes for the
honeycomb lattice, Akhmerov and Beenakker showed that, except for the
armchair edge, zigzag-type boundary condition~\cite{Brey-2006-06}, in
which two valleys decouple each other in the eigenfunctions, is
applicable for general boundary orientation with a so-called minimal
boundary, in which the edge has minimum numbers of empty sites and
dangling bonds, and these numbers are the
same.~\cite{Akhmerov-2008-02}
Above theory,~\cite{Akhmerov-2008-02} as well as assuming the slowly
varying confinement potential,~\cite{Bulaev-2008-06} supports the
decoupling of the two valley as a typical case for the m-SWNTs except
the armchair nanotubes.

Here, we will show that the two valleys are strongly coupled in the
chiral nanotubes classified into so-called metal-2 nanotubes (see \S
\ref{sec:cuttingLine}),~\cite{Saito-1998} with certain boundaries, as
well as the armchair nanotubes.
The effect of the strong coupling combined with the asymmetric
velocities appears as the vernier-scale-like spectrum.
For the so-called metal-1 nanotubes,~\cite{Saito-1998} and for the
metal-2 nanotubes with the minimal boundary, it will be shown the
fourfold degeneracy and its small lift by the spin-orbit interaction
as the result of decoupling of two valleys.
In addition, it will be shown that parity splitting of the valley
degeneracy overcomes the spin-orbit splitting for shorter metal-2
nanotubes with the minimal boundary.

In this paper, we mainly focus on the finite-length m-SWNTs in which
the ends and the center have the same rotational symmetry.
We will show that the degeneracy of discrete energy levels of m-SWNTs
strongly depends on the chirality, boundary condition, length and the
spin-orbit interaction.
We will revisit and analyse the cutting lines with the point of view
of the orbital angular momentum, then will perform numerical
calculation for an extended tight-binding model and analytical
calculation for an effective 1D model to investigate the degeneracy
and the valley coupling.

This paper is organized as follows.
In \S \ref{sec:cuttingLine}, occurrence of the valley coupling is
discussed by analyzing the cutting lines.
In \S \ref{sec:numericalCalc}, numerically calculated energy levels by
using an extended tight-binding model for metal-2 nanotubes with a
couple of boundaries are shown.
In \S \ref{sec:1Dmodel}, an effective 1D model for describing the
valley coupling is derived and the microscopic mechanism of the valley
coupling is analytically investigated for a couple of boundary
conditions.
Analytical forms of the discretized wavenumber are also given.
The conclusion is given in \S \ref{sec:Conclusion}.
In the Appendix \ref{sec:App:LongCuttingLine}, detailed analysis on
the long cutting line is given.
In the Appendix \ref{sec:App:numCalc}, numerical calculation for
metal-1, armchair and capped metal-2 nanotubes is given.
In the Appendices \ref{sec:App:Derivation1Dmodel} and
\ref{sec:App:1Dmodel_modes}, we discuss detailed derivation and mode
analysis of the 1D model, respectively.
In the Appendix \ref{sec:App:travelingMode}, relation on coefficients
of standing waves between A- and B-sublattices under a boundary is
given.

\section{Cutting line}
\label{sec:cuttingLine}

Let us consider a SWNT defined by rolling up the graphene sheet in the
direction of the chiral vector $\bm{C}_h = n \bm{a}_1 + m \bm{a}_2
\equiv (n,m)$, where $n$ and $m$ are integers specifying the chirality
of SWNT, $\bm{a}_1$ and $\bm{a}_2$ are the unit vectors of
graphene.~\cite{Saito-1998}
The m-SWNTs, which satisfy ${\rm mod} (2n + m, 3) = 0$, are further
classified into metal-1 ($d_R = d$) or metal-2 ($d_R = 3 d$), where
$d={\rm gcd}(n,m)$ is the greatest common divisor (gcd) of $n$ and
$m$, $d_R={\rm gcd}(2n+m,2m+n)$, and it has been known that the $K$
and $K'$ points sit on the center of the 1D BZ for the metal-1
nanotubes, while they sit on $1/6$ and $5/6$ positions for the metal-2
nanotubes.~\cite{Saito-1998}

In this section, we will show that the two valleys have different
orbital angular momenta for the metal-1 nanotubes, whereas they have
the same orbital angular momentum for the metal-2 nanotubes from
analysis of the cutting line, 1D BZ plotted in 2D
$k$-space.~\cite{Samsonidze:2003-12-01T00:00:00:1533-4880:431}
Here the orbital angular momentum of the valley is that at the valley
center [$K$ ($K'$) point], and is given by an integer specifying the
cutting line passing through the $K$ ($K'$) point.
The corresponding properties have been shown in the previous work
numerically.~\cite{PhysRevB.71.125408}
Here, we will show a proof of this property analytically.

States on a cutting line represent 1D wavevectors in the direction of
nanotube axis with an orbital angular momentum with respect to the
nanotube axis which corresponds to a wavevector in the circumference
direction of the SWNT.
For a finite-length SWNT, the 1D wavevectors are no longer good quantum
numbers.
If the boundaries have the same $C_d$ rotational symmetry around the
nanotube axis with that of the SWNT, the orbital angular momentum
specified by a cutting line is a conserved quantity.
In this case, an electron with a wavevector is scattered at the
boundary to 1D states with the same orbital angular momentum, that is,
the scattering within the cutting line.

\begin{figure}[htb]
  \includegraphics[width=6cm]{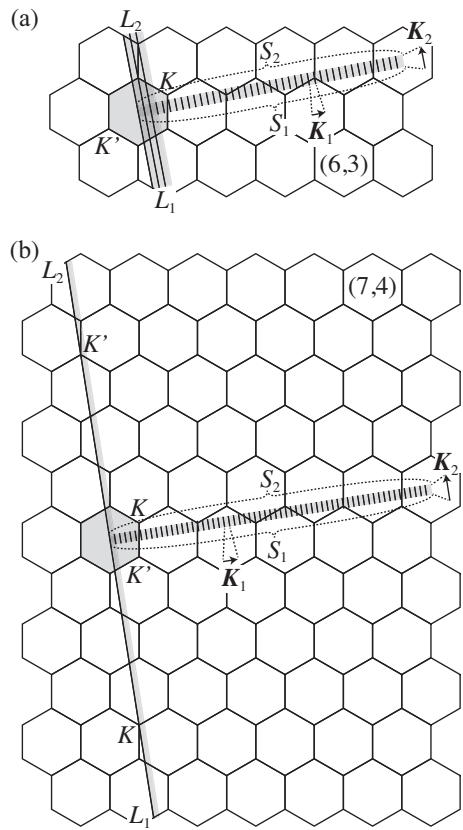}
  \caption{Cutting lines for (a) $(n,m)=(6,3)$, and (b) $(7,4)$ SWNTs.
    In each figure, shadow areas show three different choices of the
    2D BZ of graphene.  $S_1 S_2$ short segments denote conventional
    cutting lines, while $L_1 L_2$ long segment is another definition
    of 1D BZ.  Here $d=3$, $d_R=3$, $T=\sqrt{21}a$,
    $|\bm{C}_h|=3\sqrt{7}a$ and $N=42$ for $(6,3)$ SWNT, and $d=1$,
    $d_R=3$, $T=\sqrt{31}a$, $|\bm{C}_h|=\sqrt{93}a$ and $N=62$ for
    $(7,4)$ SWNT.}
  \label{fig:cuttingLine}
\end{figure}

Let us briefly review on the definition of the cutting lines for
discussing their orbital angular momenta.
There are arbitrary definitions for a complete set of the cutting
lines as well as there are arbitrary definitions for 2D BZ of graphene
as shown in Fig. \ref{fig:cuttingLine}.
The detailed description on the definitions of the cutting lines is
found in the review
article.~\cite{Samsonidze:2003-12-01T00:00:00:1533-4880:431}
Instead of the conventional definition of the cutting lines with short
segments,~\cite{Saito-1998} the following definition of the cutting
lines with long segments,
\begin{equation}
  k \frac{\bm{K}_2}{|\bm{K}_2|} + \mu \bm{K}_1, \label{eq:cuttingLine}
\end{equation}
with
\begin{equation}
  -\frac{\pi}{T} \frac{N}{d} \le k < \frac{\pi}{T} \frac{N}{d},
  ~~\text{and}~~ \mu = 0,\cdots, d-1,
  \label{eq:cuttingLine_region_long}
\end{equation}
which is derived from the helical and rotational
symmetries,~\cite{White-1993-03} is convenient to consider the
properties under the $C_d$ rotational symmetry.
[See the long segments $L_1 L_2$ in Fig. \ref{fig:cuttingLine} defined
  by Eqs. (\ref{eq:cuttingLine}) and
  (\ref{eq:cuttingLine_region_long}).]
Here the separation of cutting lines $\bm{K}_1 = ( - t_2 \bm{b}_1 +
t_1 \bm{b}_2 ) / N$ is perpendicular to the cutting lines and
represents the discreteness of the wavevector in the circumference
direction, and 
\begin{equation}
\bm{K}_2 = \frac{m \bm{b}_1 - n \bm{b}_2}{N},
\end{equation}
is the vector 
of short cutting lines in the conventional
definition,~\cite{Saito-1998} where $\bm{b}_1 = (2 \pi/a)
(1/\sqrt{3},1)$ and $\bm{b}_2 = (2 \pi/a) (1/\sqrt{3},-1)$ are the
reciprocal lattice vectors of graphene, $t_1$ and $t_2$ are integers
defined by $t_1 = (2 m + n) / d_R$, $t_2 = - (2 n + m) / d_R$.
$T = a \sqrt{3(n^2 + m^2 + nm)} / d_R$ is the 1D nanotube lattice
constant, $N = 2 (n^2 + m^2 + nm) / d_R$ is the number of A (B) atoms
in the nanotube 1D unit cell, $a = 2.46$\AA~is the lattice constant of
graphene.

The rectangle defined by the two vectors $d \bm{K}_1$ and $N \bm{K}_2
/ d$ which surrounds the set of the longer cutting lines (see the
vertically-long shadowed rectangle in Fig. \ref{fig:cuttingLine}) is
equivalent to the 2D BZ of graphene.
The corresponding unit vectors in the real space are the vector
$\bm{C}_h / d$, and the vector given by $\bm{R} = p_h \bm{a}_1 + q_h
\bm{a}_2$ where $p_h$ and $q_h$ satisfy the relation of $m p_h - n q_h
= d$.~\cite{PhysRevB.47.16671,Jishi-1994}
The component of $\bm{R}$ in the direction of nanotube axis is
expressed by
\begin{equation}
  a_z = \frac{T d}{N},
\end{equation}
which corresponds to shortest distance between two A (B) atoms in the
axis direction, because of the definition of
$\bm{R}$.~\cite{PhysRevB.47.16671,Jishi-1994}
Note that the inversion of the range of $k/2 \pi$ is equal to $a_z$.
Because each cutting line defined by Eqs. (\ref{eq:cuttingLine}) and
(\ref{eq:cuttingLine_region_long}) is equal to $N \bm{K}_2 / d$, all
independent $k$ states for the given angular momentum $\mu$ are
represented in a single cutting line.
Here, 
the orbital angular momentum of a state is defined by $\mu$ in
Eq. (\ref{eq:cuttingLine_region_long}).

It would be worthful to compare the present definition
[Eq. ((\ref{eq:cuttingLine_region_long}))] with conventional
definition of the cutting lines,~\cite{Saito-1998} $-\pi/T \le k <
\pi/T$, and $\mu = 0,\cdots, N-1$.
[See the set of the short segments $S_1 S_2$ in
  Fig. \ref{fig:cuttingLine}.]
In the conventional definition, every $d$ cutting lines belongs to the
same orbital angular momentum, that is, these cutting lines can mapped
onto a single longer cutting line by translations with reciprocal
lattice vectors.
Therefore, the long cutting lines is convenient to consider the
properties under the $C_d$ rotational symmetry since an orbital
angular momentum and a cutting line are one-to-one correspondence.

Hereafter we focus on the m-SWNTs, in which there are cutting lines
passing through the $K$ and $K'$ points.
As proven in the Appendix \ref{sec:App:LongCuttingLine} (and shown in
Figs. \ref{fig:cuttingLine} (a) and (b) as examples), the long cutting
line $L_1 L_2$ of $\mu=0$ passes through both $K$ and $K'$ points for
the metal-2 nanotubes, whereas cutting line passes only through $K$ or
$K'$ points for the metal-1 nanotubes.
The metal-2 nanotubes are further classified into metal-2$p$ and
metal-2$m$ by the conditions,~\cite{Saito-2005-10}
\begin{equation}
  {\rm mod} \left( \frac{m}{d}, 3 \right) 
  = \left\{
  \begin{array}{cl}
    1 & \text{~~for metal-2$p$}, \\
    2 & \text{~~for metal-2$m$}. \\
  \end{array}
  \right. \label{eq:metal2pmCond_01}
\end{equation}
It is also shown in the Appendix \ref{sec:App:LongCuttingLine} [and in
  Fig. \ref{fig:cuttingLine} (b) as an example] that the $K$ point is
located at $1/6$ ($5/6$) position and the $K'$ point is located at
$5/6$ ($1/6$) position on the long cutting line of $\mu=0$ for
metal-2$p$ (metal-2$m$) nanotubes.
Note that the positions of the $K$ and $K'$ points on the long 1D BZ
are the opposite to these on the short 1D BZ.
For the metal-1 nanotubes, the cutting lines of $\mu = \pm N / 3$
($\mu = \mp N / 3$) pass through the $K$ and $K'$, respectively, for
$d_X=2$ ($d_X=1$) where $d_X={\rm mod} [ (2n+m)/d,3) ]$, as shown in
the previous work for the conventional short cutting
lines.~\cite{Saito-2005-10}
Even though $\mu = \pm N / 3$ may exceed the range $0 \le \mu \le d -
1$, the expression is convenient because the $K$ and $K'$ points are
mapped onto the center of the long cutting lines, which corresponds to
1D wavenumber of $k=0$.
Other choices, e.g. $\mu = {\rm mod} [(2n + m)/3, d]$ and $\mu = {\rm
  mod} [(2m + n)/3, d]$ given in
Ref. ~\onlinecite{PhysRevB.71.125408}, may map the $K$ and $K'$ points
away from the $\Gamma$ point of 1D BZ [see the longer cutting lines of
  $\mu=2$ and $\mu=1$ in Fig. \ref{fig:cuttingLine} (a) for the
  $(n,m)=(6,3)$ metal-1 nanotube].

\section{Numerical Calculation}
\label{sec:numericalCalc}

The two valleys $K$ and $K'$ are decoupled for the finite-length
metal-1 nanotubes with the $C_d$ rotational symmetry, since the two
valleys belongs to states with different orbital angular momenta.
For this case, energy levels show nearly fourfold degeneracy and its
small lift by the spin-orbit interaction, as will be confirmed in
numerical calculation in the Appendix \S \ref{sec:app:numCalMetal1}.
On the other hand, the two valleys can couple for the finite-length
metal-2 nanotubes even both ends keep the $C_d$ rotational symmetry.
Here we perform numerical calculation of finite-length SWNTs to
investigate the valley coupling for the metal-2 nanotubes.
Vernier-scale-like spectrum will be shown for an orthogonal-shaped
boundary.
Nearly fourfold degeneracy and its small lift, which is {\it not} due
to the spin-orbit interaction for shorter nanotubes, will be shown for
a so-called minimal boundary.
Vernier-scale-like spectra for an armchair nanotube and a capped
metal-2 nanotube will also be shown in the Appendix \S
\ref{sec:app:vernier}.

The numerical calculation is done using the extended tight-binding
method,~\cite{Samsonidze-2004-12} in which $\pi$ and $\sigma$ orbitals
at each carbon atom is considered, and the hopping and overlap
integrals between the orbitals are evaluated from the {\it ab initio}
calculation~\cite{Porezag-1995-05} for interatomic distances of up to
10 bohr.
Since the systems we focus on are the finite-length nanotubes, the
electronic states are calculated by solving the generalized eigenvalue
problems with bases being from all orbitals in the systems.
The optimized geometrical structure given by the previous energy band
calculation~\cite{izumida-2009-06} is utilized for determining the
positions of carbon atoms.
Three-dimensional structure is taken into account in the calculation,
therefore the curvature
effects~\cite{izumida-2009-06,PhysRevB.85.165430} are automatically
included.
Spin degrees of freedom, and the atomic spin-orbit interaction $V_{\rm
  SO} = 6$ meV on each carbon atom,~\cite{izumida-2009-06} are also
taken into account.
A tiny magnetic field ($B = 10^{-6}$ T, corresponding spin splitting
is $\sim 10^{-8}$ meV) parallel to the nanotube axis is applied to
separate the two degenerate states of the Kramers pairs in the
calculation.
The tube axis ($z$) is chosen as the spin quantization axis.
In the following, two types of the boundaries are considered.
The first type of the boundary has a geometry constructed simply cut
by the plane orthogonal to the nanotube axis.
Here we call this boundary orthogonal boundary.
This boundary generally contains Klein-type terminations at which
terminated site neighbors two empty sites~\cite{Klein1994261} [see red
  site in Fig. \ref{fig:0704_diH_50nm} (a) for $(n,m)=(7,4)$].
The second type of the boundary has a geometry removing the
Klein-terminations from the orthogonal boundary [see
  Fig. \ref{fig:0704_50nm} (a) for $(n,m)=(7,4)$].
For both types, the edge has minimum number of empty sites [dashed
  circles in Figs. \ref{fig:0704_diH_50nm} (a) and \ref{fig:0704_50nm}
  (a)].
Further, the number of empty sites are the same with that of dangling
bonds for the second type.
The second type of the boundary is called minimal
boundary.~\cite{Akhmerov-2008-02}
In the numerical calculation, to eliminate the dangling bonds, each
single dangling bond at the ends is terminated by a hydrogen atom, and
the Klein-type termination is represented by two hydrogen
atoms.~\cite{PhysRevB.67.092406}
Both boundaries keep the $C_d$ rotational symmetry of the SWNTs.

\subsection{Vernier spectrum}
\label{sec:num_vernierSpect}

\begin{figure}[tbhp]
  \includegraphics[width=8cm]{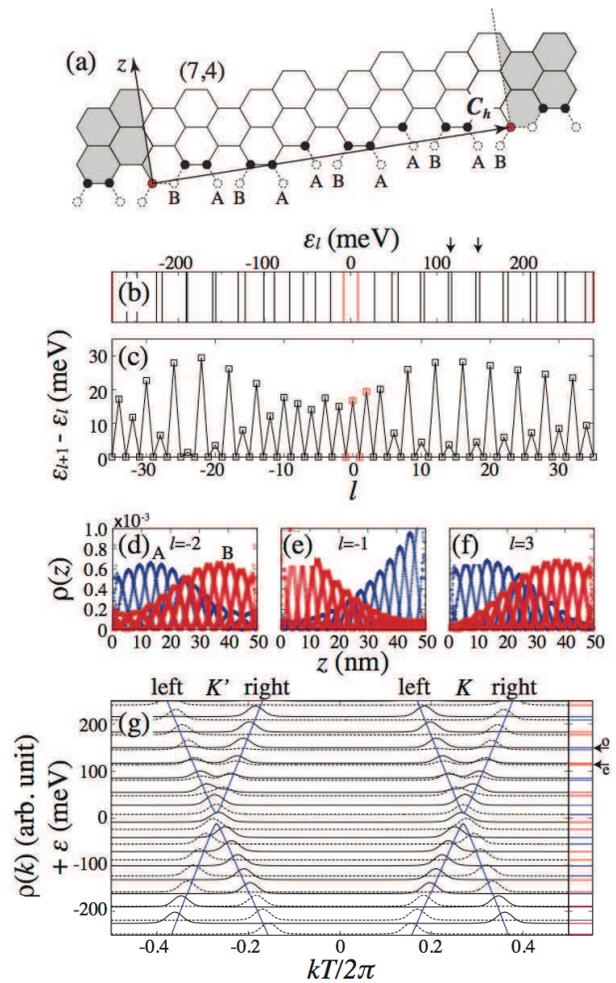}
  \caption{(Color online) Boundary shape, calculated energy levels and
    eigenstates for $(7,4)$ nanotube of $50.17$ nm length with the
    ends of orthogonal boundary.
    (a) Unfolded tube near the left end.  The empty sites are
    represented by the dashed circles, and the carbon atoms at the
    boundary are marked by the solid circles.  The red sites represent
    the Klein-type termination.  The shadowed areas repeat the
    structure in the unshadowed area.
    (b) Energy levels $\varepsilon_l$ in $-35 \le l \le 35$.  $l$ is
    the energy level index numbered in ascending order of the energy
    and $l=0$ corresponds to the HOMO level.
    (c) Level separation, $\varepsilon_{l+1} - \varepsilon_l$, as a
    function of $l$.  The levels of $-1 \le l \le 2$ indicated by red
    color in (b) and (c) are slowly decaying modes.
    (d)-(f) Local density (d) $l=-2$, (e) $l=-1$ and (f) $l=3$.  Blue
    shows that for A-sublattice, and red shows that for B-sublattice.
    (g) Zone-folded intensity plot of Fourier transform of
    wavefunction on A-sublattice for each level as a function of $k$.
    The energy for each level is added for each intensity plot.  The
    intensities for states of spin-up-majority are shown.  They are
    presented by either solid or dashed lines in turn in increasing
    the energy to show them clearly.  The blue lines show the energy
    band calculated under the periodic boundary condition.
    Right figure in (g) shows the energy levels of the even parity
    (blue lines) and the odd parity (red lines) for $V_{\rm SO}=0$.
    The arrows with e (even) and o (odd) in (g) and in (b) show the
    states exhibiting intravalley coupling with the same parity.
    In (d)-(g), the components of orbital and spin are summed up
    for each site or each wavenumber.
}
  \label{fig:0704_diH_50nm}
\end{figure}

Figure \ref{fig:0704_diH_50nm} (b) shows the calculated energy levels
$\varepsilon_l$ near the charge neutral point for (7,4) nanotube of
$50.17$ nm length with the orthogonal boundary for both ends.
Here $l$ is the energy level index numbered in ascending order of the
energy, and $l=0$ corresponds to the level of highest occupied
molecular orbital (HOMO).
To show the degeneracy behavior, the level separation,
$\varepsilon_{l+1}-\varepsilon_l$, which corresponds to the addition
energy in the tunneling spectroscopy
measurements,~\cite{Tarucha-1996-10} is plotted in
Fig. \ref{fig:0704_diH_50nm} (c).
The levels of $-1 \le l \le 2$ are slowly decaying modes that appear in
the energy gap caused by the curvature of nanotube surface [the local
  densities of $l = -2$, $l = -1$ and $l = -3$ are shown in
  Figs. \ref{fig:0704_diH_50nm} (d)-(f) as examples].
The origin of the slowly decaying modes will be discussed in \S
\ref{sec:1Dmodel}.
The level separation show oscillatory behavior between nearly fourfold
[near $l \sim -25$ ($\varepsilon \sim -190$ meV) and $l \sim 13$
  ($115$ meV)] and twofold [$l \sim -7$ ($\varepsilon \sim -55$ meV)]
degeneracies.
The behavior is understood by (i) the asymmetric velocities between
left- and right-going waves in the same valley pointed out in the
previous work,~\cite{PhysRevB.85.165430} and (ii) the strong
intervalley coupling.
The strong intervalley coupling is confirmed by the intensity plot in
the wavenumber as shown in Fig. \ref{fig:0704_diH_50nm} (g).
In this plot, the intensities for the states of spin-up-majority (with
spin-up polarization more than $50\%$, in the calculation the
polarization exceeds $99\%$) are shown.
Note that the intensity at $k$ for the spin-up-majority and that at
$-k$ for the spin-down-majority in spin degenerate levels are the same
because of the time-reversal symmetry.
For this case, the intensity plots for the spin-up-majority and that
for the spin-down-majority almost coincides with each other for each
of degenerate levels, that is, the orbital state of spin-up-majority
and that of spin-down-majority are almost the same.
Each eigenfunction shows the strong intensity only at the left-going
wave at the $K$-valley ($K'$-valley) and right-going wave at the
$K'$-valley ($K$-valley).
The strong intervalley coupling combined with the asymmetric
velocities causes the two types of the equal interval energy levels,
$\hbar v_{\rm L}^{(K)} \pi / L_{\rm NT}$ and $\hbar v_{\rm R}^{(K)}
\pi / L_{\rm NT}$, like the vernier scale.~\cite{PhysRevB.85.165430}
The period of the two- to fourfold oscillation is not constant but has
the energy dependence, for instance, the period becomes longer for the
positive energy region.
This is because the velocities has energy dependence reflecting the
deviation from the linear energy band.
The velocity difference between left- and right-going waves becomes
smaller for the higher energy region in the conduction band.
We can also see the intravalley coupling when the two levels are
close to each other, for instance, $\varepsilon_l \sim$ 83 meV and 115
meV as shown by arrows in the left of Fig. \ref{fig:0704_diH_50nm}
(g).
The coupling occurs between the same parity states which will be
discussed in \S \ref{sec:1Dmodel}.

\subsection{Valley degeneracy and lift of degeneracy}
\label{sec:num_4fold}

\begin{figure}[hbtp]
  \includegraphics[width=8cm]{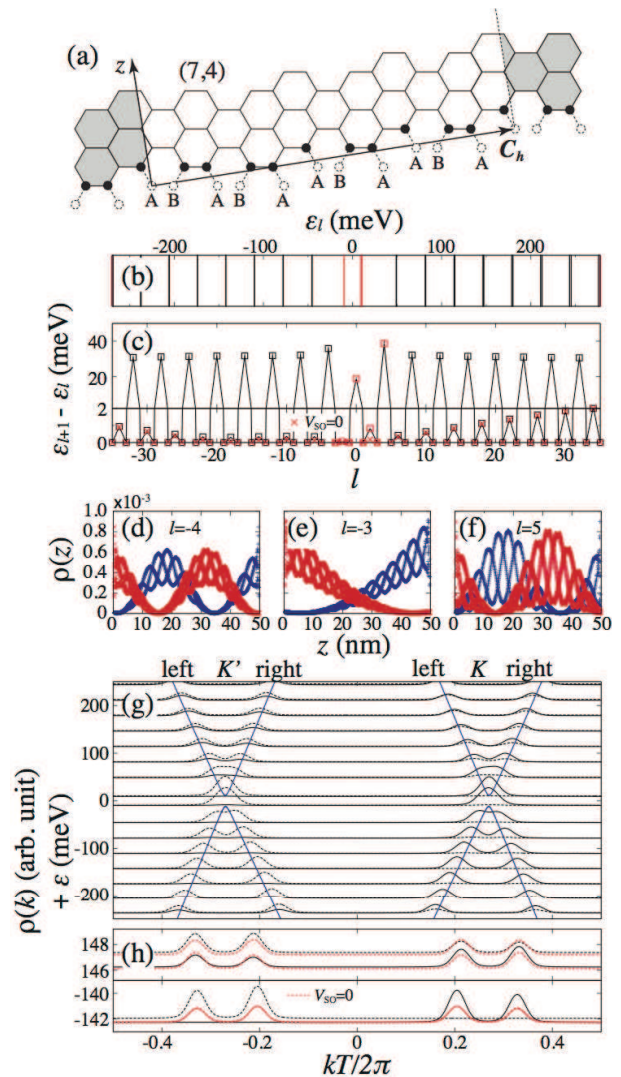}
  \caption{(Color online) Boundary shape, calculated energy levels and
    eigenstates for $(7,4)$ nanotube of $50.17$ nm length with the
    ends of minimal boundary.
    (a) Unfolded tube near the left end.
    (b) Energy levels $\varepsilon_{l}$ in $-35 \le l \le 35$.
    (c) Level separation $\varepsilon_{l+1} - \varepsilon_l$ as a
    function of $l$.  The case of absence of spin-orbit interaction is
    shown by the red cross in the lower panel.
    (d)-(f) Local density for (d) $l=-4$, (e) $l=-3$ and (f)
    $l=5$.
    (g) The zone-folded intensity plot of Fourier transform of
    wavefunction on A-sublattice for each level.  The intensities for
    the states of spin-up-majority are shown.
    (h) Enlarged plot of (g) near $\varepsilon_l = 147$ meV (upper
    panel) and $\varepsilon_l = -142$ meV (lower panel).  The case of
    absence of spin-orbit interaction is shown by the red dashed
    lines.
  }
  \label{fig:0704_50nm}
\end{figure}

Figure \ref{fig:0704_50nm} (b) shows the calculated energy levels
$\varepsilon_l$ for $50.17$ nm length (7,4) nanotube with the minimal
boundary for both ends.
The levels of $-3 \le l \le 4$ indicated with red lines are the slowly
decaying modes in the energy gap [the local density of $l = -3$ is
  shown in Fig. \ref{fig:0704_50nm} (e) as an example].
Above and below the energy gap, the level separation shows almost
equal interval reflecting the linear energy dispersion.
Fig. \ref{fig:0704_50nm} (c) shows that each of levels shows nearly
fourfold degeneracy, which is very different from that in
Fig. \ref{fig:0704_diH_50nm} (c).
As shown in the lower panel of Fig. \ref{fig:0704_50nm} (c), the
degeneracy is lifted of the order of meV.
The lift of the degeneracy is clearly observed even for the absence of
the spin-orbit interaction in some energy region, for instance,
$\varepsilon \gsim 100$ meV.
From the intensity plot in the wavenumber in Fig. \ref{fig:0704_50nm}
(g), each level show almost equal intensity between the left- and
right-going waves in the same valley, showing the valley degeneracy.
The enlarged plot near $\varepsilon_l = 147$ meV and $\varepsilon_l =
-142$ meV in Fig. \ref{fig:0704_50nm} (h) shows the effect of the
spin-orbit interaction.
The four levels near $\varepsilon_l = 147$ meV show splitting into two
spin-degenerate levels with $\Delta \varepsilon \sim 2$ meV when the
spin-orbit interaction is absent.
The splitting and the intensities does not change much when the
spin-orbit interaction is turned on.
The four levels near $\varepsilon_l = -142$ meV, on the other hand,
show almost degeneracy when the spin-orbit interaction is absent.
When the spin-orbit interaction is turned on, they splits into two
spin-degenerate levels with the splitting $\Delta \varepsilon \sim
0.3$ meV.
At the same time, the state of spin-up-majority in lower energy shows
the intensity only at the $K$-valley ($kT/2\pi \sim 0.3$) whereas that
in higher energy shows the intensity only at the $K'$-valley ($kT/2\pi
\sim -0.3$).
The splitting is thus induced by both an intrinsic property of
finite-length nanotubes with the minimal boundary and the spin-orbit
interaction.

\begin{figure}[htb]
  \includegraphics[width=8cm]{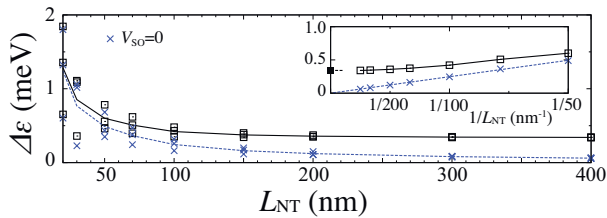}
  \caption{(Color online) Length dependence of splitting of fourfold
    degeneracy for $(7,4)$ nanotube with minimal boundary.
    Three same symbols at each length show the averaged splitting in
    the energy range $|\varepsilon_l| < 250$ meV except the evanescent
    modes.
    The splittings are then also averaged for the three cases of the
    adjacent lengths, $L_{\rm NT}$, $L_{\rm NT} + a_z$, $L_{\rm NT} +
    2a_z$, because the splitting shows nearly three-fold oscillations
    with respect to the nanotube length in the unit of $a_z = 0.022$
    nm.
    Solid line shows the averaged splitting with the spin-orbit
    interaction, $V_{\rm SO} = 6$ meV, dashed line shows that without
    the spin-orbit interaction.
    Inset shows the splitting as a function of the inverse of the
    length.  The solid square with a bar (at $0.34$ meV) shows the
    absolute value of the spin-splitting averaged in the energy range
    $|\varepsilon_l| < 250$ meV calculated from the energy band
    calculation.
}
  \label{fig:DelE_fourfold-L_0704}
\end{figure}

Fig. \ref{fig:DelE_fourfold-L_0704} shows the splitting of the
fourfold degeneracy for $(7,4)$ nanotube with minimal boundary as a
function of nanotube length.
The splitting is averaged in the energy range $|\varepsilon_l| < 250$
meV except the evanescent modes.
Further, because the splitting shows nearly three-fold oscillations
with respect to the nanotube length in the unit of $a_z$, the average
for the three cases is also taken.
Without the spin-orbit interaction, the averaged splitting decreases
when the length increases with $1/L_{\rm NT}$ dependence.
For the actual case of the presence of spin-orbit interaction, the
splitting for the longer nanotubes ($\gsim 200$ nm) are dominated by
the spin-orbit interaction, as shown that the splitting is asymptotic
to a constant value ($\sim 0.34$ meV).

\section{Effective 1D model}
\label{sec:1Dmodel}

The numerical calculation showed that the degeneracy of the energy
levels, and the valley coupling in the eigenfunctions, are quite
sensitive to the boundary shape.
To investigate the microscopic mechanism of the valley coupling for
given boundary conditions, we will introduce an effective 1D model,
which extract the states near the two valleys remaining the atomic
structure.
This model can treat microscopic analysis of the coupling of two
valleys, which can be an advantage from the conventional effective
mass theory which treats each valley separately.
It will be shown that the model shows evanescent and traveling nature
of wavefunctions, and the comparison of the numbers of evanescent
modes and the boundary conditions at the ends is an important key to
understand the behaviors of valley coupling in the standing waves
composed from the traveling modes.
Analytical form of the discrete wavenumbers, the length dependence of
the degeneracy lift, and the slowly decaying modes in the energy gap
will also be shown.

\subsection{Effective 1D model}
\label{sec:1DmodelDerivation}

Let us consider the following nearest-neighbor tight-binding
Hamiltonian,
\begin{equation}
  H = \sum_{\vec{r}} \sum_{j=1}^3 \gamma_j c_{{\rm A}, \vec{r}}^\dagger c_{{\rm B}, \vec{r} + \vec{\Delta}_j} 
  + {\rm H.c.}, \label{eq:H_00}
\end{equation}
where $c_{{\rm A}, \vec{r}}^\dagger$ is the creation operator of $\pi$
electron on A atom at site $\vec{r}$, and $c_{{\rm B}, \vec{r} +
  \vec{\Delta}_j}$ is the annihilation operator of $\pi$ electron on
the neighbor $j$-th B atom ($j=1,2,3$) at $\vec{r} + \vec{\Delta}_j$,
$\vec{\Delta}_j$ is the vectors from A to nearest $j$-th B atoms (see
$\vec{\Delta}_j$ in Fig. \ref{fig:Coordinates_0704}).
The summation on $\vec{r}$ is taken over the finite-length SWNTs.
$\gamma_j$ is the hopping integral between A and $j$-th B atom which
is chosen as the real number, and, in general, they are different from
one another because of the curvature of nanotube surface.
For the simplicity, we ignore the spin degrees of freedom, the
$\sigma$-orbitals, and the hopping to next nearest neighbor and
further sites.
The valley coupling in the standing waves can be discussed within this
simplified model as shown later.
The asymmetric velocities~\cite{PhysRevB.85.165430} and the spin-orbit
interaction~\cite{izumida-2009-06} are not able to be captured in this
model, and they affect as the vernier-like spectrum and the spin-orbit
splitting as shown in the numerical calculation in \S
\ref{sec:numericalCalc}.

\begin{figure}[htb]
  \includegraphics[width=8cm]{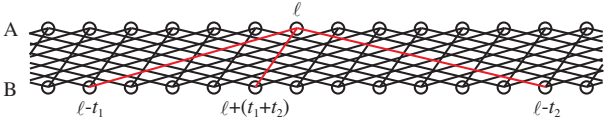}
  \caption{(Color online) Effective one-dimensional model for
    $(n/d,m/d)=(7,4)$ SWNTs in which $(t_1,t_2)=(5,-6)$.  The solid
    lines show the hopping between atoms.  The red lines denote the
    hopping from (and to) A atom at $\ell$-site.}
  \label{fig:1Dmodel_0704}
\end{figure}

The construction of an effective 1D model is performed by projecting
to an angular momentum $\mu$, as previously done for the achiral
nanotubes.~\cite{PhysRevB.55.R11973}
The effective 1D Hamiltonian for the metal-2 nanotubes is given by
(see the Appendix \ref{sec:App:Derivation1Dmodel} for the derivation),
\begin{equation}
  H_{\mu=0}
  = \sum_\ell \sum_{j=1}^3 \gamma_j a_{\ell}^\dagger b_{\ell + \Delta \ell_j} + {\rm H.c.}, 
  \label{eq:H1D_metal2}
\end{equation}
where 
\begin{equation}
  \Delta \ell_1 =   t_1 + t_2, ~~~~
  \Delta \ell_2 = - t_2, ~~~~
  \Delta \ell_3 = - t_1,       \label{eq:Deltaell_j}
\end{equation}
and 
\begin{equation}
  a_{\ell} = 
  \frac{1}{\sqrt{d}} 
  \sum_{\vec{r}} c_{{\rm A}, {\vec{r}}} \delta_{a_z \ell, z_{\vec{r}}}, ~~~~
  b_{\ell} = 
  \frac{1}{\sqrt{d}} 
  \sum_{\vec{r}} c_{{\rm B}, {\vec{r}}} \delta_{a_z \ell, z_{\vec{r}}},
\end{equation}
correspond to the $\mu=0$ Fourier components of the operators.
The index $\ell$ is an integer specifying the 1D site.
The summation is taken over for $d$ A atoms of $d$ B atoms which
satisfy $z_{\vec{r}} = a_z \ell$ for $a_{\ell}$ and $b_{\ell}$,
respectively.
Note that there is a pair of A and B atoms on each $\ell$ because
there are $d$ pairs of A and B atoms on the same $z$ for the metal-2
nanotubes.
The model for $(n/d,m/d)=(7,4)$ SWNTs is depicted in
Fig. \ref{fig:1Dmodel_0704}.
In Fig. \ref{fig:1Dmodel_0704}, A$_\ell$ atom is connected to
B$_{\ell+6}$, B$_{\ell-1}$, B$_{\ell-5}$ atoms.
Note that the Hamiltonian $H_{\mu=0}$ for $(7,4)$ SWNT is the total
Hamiltonian since $d=1$.
For this case the effective model has the same bond connection with
the original model.

\subsection{Modes of 1D model}
\label{sec:1Dmodel_modes}

Eigenfunctions of Eq. (\ref{eq:H1D_metal2}) are expressed as linear
combinations of independent modes of the Hamiltonian.
The modes of Eq. (\ref{eq:H1D_metal2}) are classified into traveling
modes and evanescent modes.
Coefficients of the traveling modes (and the evanescent modes) in each
eigenfunction are determined by boundary conditions as will be
discussed in the subsections below (\S \ref{sec:1Dmodel_finiteLength}
- \ref{sec:boundaryCondition}).
Here we will show the independent modes of the Hamiltonian.

Hereafter we will consider the cases of $n \ge m > 0$ for the metal-2
nanotubes, and then we have $|t_2| = - t_2 > 0$ and $t_1 > 0$.
For an eigenstate $| \Phi \rangle = \sum_{\sigma \ell} \phi_{\sigma
  \ell} | \sigma \ell \rangle$ with energy $\varepsilon$ of the
Hamiltonian (\ref{eq:H1D_metal2}), where $| \sigma \ell \rangle$ is
the $\pi$-state at $\sigma$-atom ($\sigma={\rm A, B}$) on $\ell$-site,
we have the following equations of motion,
\begin{align}
  & \gamma_1 \phi_{{\rm A} \ell + (|t_2| - t_1)}
  + \gamma_2 \phi_{{\rm A} \ell - |t_2|}
  + \gamma_3 \phi_{{\rm A} \ell + t_1} = \varepsilon \phi_{{\rm B} \ell}, \label{eq:EOM_A} \\
  & \gamma_1 \phi_{{\rm B} \ell - (|t_2| - t_1)}
  + \gamma_2 \phi_{{\rm B} \ell + |t_2|}
  + \gamma_3 \phi_{{\rm B} \ell - t_1} = \varepsilon \phi_{{\rm A} \ell}. \label{eq:EOM_B}
\end{align}
Substituting the following forms for the solutions,
\begin{equation}
  \phi_{\sigma \ell + t} = \lambda^t \phi_{\sigma \ell}, ~~~~
  \phi_{{\rm B} \ell} = \eta \phi_{{\rm A} \ell}, \label{eq:EOM_Sol}
\end{equation}
for Eqs. (\ref{eq:EOM_A}) and (\ref{eq:EOM_B}), one gets the following
simultaneous equations,
\begin{align}
  & \gamma_1 \lambda^{|t_2| - t_1} + \gamma_2 \lambda^{-|t_2|}  + \gamma_3 \lambda^{t_1}  
  = \varepsilon \eta,         \label{eq:EOM_A_alg} \\
  & \gamma_1 \lambda^{- (|t_2| - t_1)}    + \gamma_2 \lambda^{|t_2|} + \gamma_3 \lambda^{-t_1} 
  = \frac{\varepsilon}{\eta}. \label{eq:EOM_B_alg}
\end{align}
Since $t_1 + |t_2| = n/d + m/d$, there are $2(n/d + m/d)$ sets of
solutions $(\lambda, \eta)$ for Eqs. (\ref{eq:EOM_A_alg}) and
(\ref{eq:EOM_B_alg}) because each equation is the $(n/d + m/d)$-th
order algebraic equation.
For the case of $|\eta| < 1$, we can call the mode A-like mode because
the wavefunction is polarized at A atoms.
On the other hand, we can call B-like mode for $|\eta| > 1$.
Further, we can call evanescent mode at left (right) side for $|
\lambda | < 1$ $( > 1)$.
The mode with $|\lambda| = |\eta| = 1$ is called traveling mode.
Let $\gamma$ be the average of $\gamma_j$, $\gamma = \sum \gamma_j /
3$, and $\delta\gamma_j$ be the difference from the average,
$\delta\gamma_j = \gamma_j - \gamma$.
Because of the curvature of nanotube surface, $\delta \gamma_j \ne 0$.
Hereafter we restrict our situation to consider the low energy states,
$|\varepsilon / \gamma | \ll 1$, and small deviation of the hopping
integrals from the average value, $|\delta\gamma_j / \gamma | \ll 1$.
Hereafter of this subsection we will show main results for the total
$2(n/d + m/d)$ modes.
The detail derivation of the modes are given in Appendix
\ref{sec:App:1Dmodel_modes}.

It is shown that there are $(n/d + m/d - 2)$ A-like modes and $(n/d +
m/d - 2)$ B-like modes in the energy range considered.
Further, by following Appendix B of
Ref. ~\onlinecite{Akhmerov-2008-02}, the A-like modes are classified
into $|t_2| - 1$ evanescent modes at left side and $t_1 - 1$
evanescent modes at right side, whereas B-like modes are classified
into $t_1 - 1$ evanescent modes at left side and $|t_2| - 1$
evanescent modes at right side.

The remaining four modes are the traveling modes, or slowly decaying
evanescent modes, depending on the energy.
For the energy outside the energy gap induced by the curvature of
nanotube
surface,~\cite{Hamada-1992-03,Saito-1992-PRB,Kane-1997-03,Ando-2000-06,izumida-2009-06}
$| \varepsilon | > \varepsilon_{\rm gap}$, there are the following
four traveling modes,
\begin{equation}
  (\lambda, \eta) = \left( e^{{\rm i} k_\ell}, e^{{\rm i} \Phi(k_\ell)} \right), ~~~~
  k_\ell = \tau' k_0 + k, \label{eq:travelingModes_gap}
\end{equation}
for the energies
\begin{equation}
  \varepsilon = \pm |\gamma| \frac{\sqrt{3} a}{2 a_z} \sqrt{ k^2 + k_{\rm I}^2 }. 
  \label{eq:energy_k_gap}
\end{equation}
Here $k_\ell$ in Eq. (\ref{eq:travelingModes_gap}) denotes four
wavenumbers, where $\tau' = \pm 1$ indicates the two valleys, and $k$,
a wavenumber measured from $\tau' k_0$, becomes either positive or
negative, where $k_0 = 2 \pi / 3 + k_{\rm R}$.
[Note that $\tau = - \tau' \beta$ corresponds to the index for the $K$
  ($\tau = 1$) or $K'$ ($\tau=-1$) points, where $\beta = 1$ ($\beta =
  -1$) is introduced for the metal-2$p$ (metal-2$m$) nanotubes.]
$k_{\rm R}$ and $k_{\rm I}$ are the shift of the Dirac point in
$\bm{K}_2$ and $\bm{K}_1$ direction, respectively, at the $\tau' = 1$
valley, because of the curvature of nanotube surface.
From the previous energy band calculation with the extended
tight-binding method,~\cite{izumida-2009-06} they are evaluated to be
\begin{equation}
  k_{\rm R} = -a_z \beta \zeta \frac{\sin 3 \theta}{d_t^2}, 
  ~~~~ 
  k_{\rm I} = a_z \beta' \frac{\cos 3 \theta}{d_t^2}, \label{eq:kR_kI_compare_bandCalc}
\end{equation}
where $d_t = |\bm{C}_h| / \pi$ is the diameter of nanotube, $\theta =
\arccos (2n + m)/2\sqrt{n^2 + m^2 + nm}$ is the chiral angle, and the
coefficients are evaluated to be $\beta' = 0.0436$ nm and $\zeta =
-0.185$ nm.
The energy dispersion of Eq. (\ref{eq:energy_k_gap}) shows the energy
gap
\begin{equation}
  \varepsilon_{\rm gap} 
  =
  |\gamma| \frac{\sqrt{3} a}{2 a_z} |k_{\rm I}|.
\end{equation}
The phase $\Phi$ in Eq. (\ref{eq:travelingModes_gap}) is given by
\begin{equation}
  \Phi(\tau' k_0 + k) 
  = \tau' \left( \frac{2 \pi}{3} t_2 + \beta \theta \right)
  + \arg \left[ \frac{\gamma \left( k + {\rm i} k_{\rm I} \right) }{{\rm i} \varepsilon} \right].
  \label{eq:phase_gap}
\end{equation}

Inside the gap, $|\varepsilon| < \varepsilon_{\rm gap}$, there are no
traveling modes, but four slowly decaying evanescent modes, $( e^{i
  \tau' k_0 - \kappa}, \eta )$, exist for the energies
\begin{equation}
  \varepsilon = \pm |\gamma| \frac{\sqrt{3} a}{2 a_z} \sqrt{ k_{\rm I}^2 -\kappa^2 }. 
  \label{eq:energy_InsideGap}
\end{equation}
Note that $\kappa$ can be either positive or negative.
We have
\begin{equation}
  | \eta | = \frac{ | k_{\rm I} + \kappa | }{ \sqrt{ k_{\rm I}^2 - \kappa^2 }  }.
\end{equation}
For the case of $k_{\rm I} > 0$, $|\eta| = \sqrt{ \left( k_{\rm I} +
  \kappa \right) / \left( k_{\rm I} - \kappa \right) }$, then there
are two B-like modes ($|\eta| > 1$) which are slowly decaying near the
left end ($\kappa > 0$), and two A-like modes ($|\eta| < 1$) which are
slowly decaying modes near the right end ($\kappa < 0$).
The decay length is estimated to be $\sim 1/|k_{\rm I}|$, that could
be much longer than that for the other $2(n/d + m/d - 2)$ evanescent
modes having the decay length of the order of carbon-carbon bond
length.
The slowly decaying modes appeared in the numerical calculation as
shown in Fig. \ref{fig:0704_diH_50nm} (e) and Fig. \ref{fig:0704_50nm}
(e).
Note that the appearance of slowly decaying modes for each sublattice
depends on the sign of $k_{\rm I}$.
Within the nearest-neighbor tight-binding model, the value might be
estimated to be positive or negative with a small value depending on
the application of the hopping parameters.~\cite{Ando-2000-06}
If $k_{\rm I}$ is estimated to be a negative value, the slowly
decaying modes could appear by applying the Aharonov-Bohm (AB)
flux.~\cite{PhysRevB.71.195401,PhysRevB.77.045138,PhysRevB.83.193407}
We found that $k_{\rm I}$ is a positive value in the previous extended
tight-binding calculation,~\cite{izumida-2009-06} and the present
numerical calculation showing the slowly decaying modes consistent
with $k_{\rm I} > 0$.

\subsection{Eigenfunctions of finite-length 1D model}
\label{sec:1Dmodel_finiteLength}

An eigenstate of the finite-length 1D model is expressed by a linear
combination of the modes derived in \S \ref{sec:1Dmodel_modes}.
Let us extract only the traveling modes given in
Eq. (\ref{eq:travelingModes_gap}) in the eigenstate.
Because of the real number of the hopping integrals in the Hamiltonian
with $\mu=0$, the components of the traveling modes in the eigenstates
are written by real functions of standing waves,
\begin{align}
  \phi_{{\rm A} \ell} 
  = \sum_{r = \pm 1} 
  c_r \cos \left[ \left( k_0 + k_r \right) \ell - \frac{\Phi (k_0 + k_r)}{2} + \theta_r \right],
  \label{eq:standingWave_A} \\
  \phi_{{\rm B} \ell} 
  = \sum_{r = \pm 1} 
  c_r \cos \left[ \left( k_0 + k_r \right) \ell + \frac{\Phi (k_0 + k_r)}{2} + \theta_r \right],
    \label{eq:standingWave_B}
\end{align}
where $r = +1$ ($r = -1$) denotes the branch of right-going
(left-going) wave at $k_\ell = k_0$.
$k_r$ is the wavenumber of left-going or right-going waves measured
from $k_\ell = k_0$.
The cosine function for $r=+1$ ($r=-1$) is the linear combination of
the right-going (left-going) wave at $k_\ell = k_0$ and its
time-reversal state of the left-going (right-going) wave at $k_\ell =
-k_0$.
For the case that the left and right-going waves have the same
velocity, the relation $k_{r=-1} = - k_{r=+1}$ holds in
Eqs. (\ref{eq:standingWave_A}) and (\ref{eq:standingWave_B}) because
the left- and right-going waves which compose $\phi_{{\rm A} \ell}$
and $\phi_{{\rm B} \ell}$ should have the same energy.
For the asymmetric velocity case,~\cite{PhysRevB.85.165430} however,
they are different from each other.
We can choose that the coefficients to be positive, $c_r \ge 0$, and
the phase $\theta_r$ to be real numbers.
The coefficients, the phases, and the discretized wavenumbers $k_r$
are determined by applying boundary conditions at the left and the
right ends.

It should be noted that the functions of
Eqs. (\ref{eq:standingWave_A}) and (\ref{eq:standingWave_B})
themselves do not always describe the strong intervalley coupling,
even though they have the functional form of linear combination of two
valley states.
For instance, when the two valleys are completely decoupled, each
valley state as a linear combination of left- and right-going wave in
the same valley is an eigenstate.
Since the two valley states are degenerate, a linear combination of
the two states, $c_{+1} = c_{-1}$ in Eqs. (\ref{eq:standingWave_A})
and (\ref{eq:standingWave_B}), is also an eigenstate of the
Hamiltonian.
When the spin-orbit interaction is introduced, the splitting of the
doubly degenerate energy band occurs because of the lack of the
inversion symmetry for the chiral
nanotubes.~\cite{Chico-2004-10,izumida-2009-06}
The valley degeneracy is lifted by the spin-orbit interaction, and the
$K$-valley state with spin-up (spin-down) function and the $K'$-valley
state with spin-down (spin-up) function are the set of
eigenfunctions for each Kramers pairs, as shown in the negative energy
region in Figs. \ref{fig:0704_50nm} (g) and (h).
For this case, Eqs. (\ref{eq:standingWave_A}) and
(\ref{eq:standingWave_B}) do not describe the orbital states for each
spin state.
However, for cases that the spin-orbit interaction is irrelevant such
as Fig. \ref{fig:0704_diH_50nm} and $\varepsilon_l \gsim 100$ meV
region in Fig. \ref{fig:0704_50nm}, Eqs. (\ref{eq:standingWave_A}) and
(\ref{eq:standingWave_B}) could express the approximated orbital parts
of the eigenstates in the finite-length metal-2 nanotubes.

\subsection{Parity symmetry}
\label{sec:1Dmodel_parity}

In order to discuss the boundary conditions, let us consider the parity
symmetry of the 1D model, that is, the Hamiltonian is invariance by
exchanging $a_\ell \leftrightarrow b_{N_s + 1 - \ell}$, where $N_s$ is
the site index at the right end.
This corresponds to an inversion symmetry of 1D lattice of
Fig. \ref{fig:1Dmodel_0704}.
Using the parity symmetry, it is enough to consider only one of the
two ends instead of both ends for the boundary conditions.
We have the following relation on the wavefunctions,
\begin{equation}
  \phi_{{\rm A} \ell} = p \phi_{{\rm B} N_s + 1 - \ell} ~~\text{for any}~\ell, \label{eq:parityRelation}
\end{equation}
where $p = \pm 1$ are the parity eigenvalues.
Applying Eq. (\ref{eq:parityRelation}) to the standing waves in
Eqs. (\ref{eq:standingWave_A}) and (\ref{eq:standingWave_B}), it is
shown that one of the following three conditions should be satisfied,
\begin{equation}
  \left\{
  \begin{array}{l}
    c_{+1} = 0, \\
    \exp 
    \left( {\rm i} \left[ \left( k_0 + k_{-1} \right) \left( N_s + 1 \right) + 2 \theta_{-1} \right] \right)
      = p, \\
  \end{array}
  \right.
  \label{eq:from_parity_01}
\end{equation}
or
\begin{equation}
  \left\{
  \begin{array}{l}
    c_{-1} = 0, \\
    \exp 
    \left( {\rm i} \left[ \left( k_0 + k_{+1} \right) \left( N_s + 1 \right) + 2 \theta_{+1} \right] \right)
    = p, \\
  \end{array}
  \right.
  \label{eq:from_parity_02}
\end{equation}
or
\begin{align}
  \left\{
  \begin{array}{l}
    c_{r} \ne 0, \\
    \exp 
    \left( {\rm i} \left[ \left( k_0 + k_r \right) \left( N_s + 1 \right) + 2 \theta_{r} \right] \right)
    = p, \\
  \end{array}
  \right.   &   \label{eq:from_parity_03} \\
  \text{for both } r = \pm 1. & \nonumber 
\end{align}
The selection from Eqs. (\ref{eq:from_parity_01}) -
(\ref{eq:from_parity_03}) depends on the boundary type, as is
discussed below.
Note that the parity symmetry of the 1D model corresponds to the $C_2$
rotational symmetry around the $C_2$ axis at center of a carbon-carbon
bond in the direction perpendicular to the nanotube axis in the
original SWNT.~\cite{Barros-2006-09}
This is {\it not} an inversion symmetry of the original SWNT.
Unlike the inversion operation, the $C_2$ rotation change the
spin-direction.
The parity symmetry for the eigenfunctions is broken in the presence
of the spin-orbit interaction since the orbital state and the spin
state are coupled.
This is the reason why the parity states for the absence of the
spin-orbit interaction are shown in Fig. \ref{fig:0704_diH_50nm} (g).

\subsection{Boundary conditions}
\label{sec:boundaryCondition}

\begin{figure}[htb]
  \includegraphics[width=8cm]{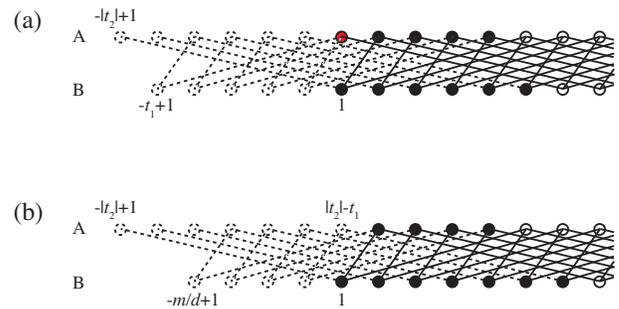}
  \caption{(Color online) Two examples of the left end classified into
    (a) orthogonal boundary condition in which $|t_2|$ A sites and
    $t_1$ B sites are empty, and (b) minimal boundary condition in
    which $n/d$ A sites and $m/d$ B sites are empty, for $(n/d,
    m/d)=(7,4)$ nanotubes.  The dashed circles represent the empty
    atomic sites and the dashed lines represent the missing bonds.
    The solid circles represent the carbon atoms at the boundary, and
    the red circle represents the Klein-type termination.}
  \label{fig:1D_boundary_0704}
\end{figure}

There is a variety on the geometry of the boundary shape.
Here we explicitly consider two types of boundary, one gives strong
valley coupling and another gives decoupling of two valleys.
For a moment we exclude the armchair nanotubes which are classified
into metal-2 nanotubes.
The first type of the boundary is depicted in
Fig. \ref{fig:1D_boundary_0704} (a).
At the left end, both A and B atoms are terminated at the same $z$
coordinate (at $\ell=1$).
The end is constructed by cutting nanotube orthogonal to the axis
direction.
This is the orthogonal boundary.
Each A atom in $1 \le \ell \le |t_2| - t_1$ connects to a B atom,
corresponds to the Klein-type termination.~\cite{Klein1994261}
Each A atom in $|t_2| - t_1 + 1 \le \ell \le t_1$ connects to two B
atoms, and each B atom in $1 \le \ell \le |t_2|$ connects to two A
atoms.
For this case, the following boundary conditions are imposed,
\begin{align}
  &  \phi_{{\rm A} \ell} = 0 
  ~~\text{at}~~ \ell = - |t_2| + 1, \cdots, 0. 
  \label{eq:BC_perp_A} \\
  &  \phi_{{\rm B} \ell} = 0
 ~~\text{at}~~ \ell = -  t_1  + 1, \cdots, 0. 
  \label{eq:BC_perp_B}
\end{align}
The number of boundary conditions for $\phi_{{\rm A} \ell}$ is
$|t_2|$, and that for $\phi_{{\rm B} \ell}$ is $t_1$.
As shown in \S \ref{sec:1Dmodel_modes}, in the low energy region,
there are totally $|t_2| + 1$ relevant modes (two traveling and $|t_2|
- 1$ evanescent modes) for $\phi_{{\rm A} \ell}$, and $t_1 + 1$
relevant modes for $\phi_{{\rm B} \ell}$ at the left end.
Therefore, the wavefunctions for A- and B-sublattices can be
determined with an arbitrarity of the amplitude.
In addition, as shown in the Appendix \ref{sec:App:travelingMode}, the
wavefunctions of A- and B-sublattices share a common coefficient for
this case.
Therefore, we have two solutions for the wavefunctions of
Eqs. (\ref{eq:standingWave_A}) and (\ref{eq:standingWave_B}):
$c_{+1}=0$ and $\theta_{-1}$ is determined, or, $c_{-1}=0$ and
$\theta_{+1}$ is determined.
The coefficient of $c_{+1} = 0$ or $c_{-1} = 0$ reflects only the
intervalley scattering at the ends.
For this case, either Eq. (\ref{eq:from_parity_01}) or
Eq. (\ref{eq:from_parity_02}) should be satisfied.
Therefore, we get the following discretization for the wavenumbers,
\begin{equation}
  k_r = \frac{ l_p \pi }{N_s + 1} + \delta k_r, \label{eq:k_disc_01}
\end{equation}
where $l_p$ is an even (odd) integer for $p=1$ ($p=-1$), 
$\delta k_r = (2 L \pi - 2 \theta_r) / (N_s + 1) - k_0$ 
is a small offset, an integer $L$ may be chosen as 
$L = [ 2 \theta_r + k_0 (N_s + 1) ] / 2 \pi $, 
where $[x]$ is Gauss's symbol representing the greatest integer that
is less than or equal to $x$.
In general, $\delta k_{r=+1} \ne \delta k_{r=-1}$, therefore the
energy levels are not degenerate between $r = +1$ and $r = -1$.
The corresponding expression is used in the previous
articles~\cite{Rubio-1999-04,Yaguchi-2001-05,Mayrhofer-2006-09,Mayrhofer-2007,PhysRevB.84.165427}
for the armchair nanotubes.
Eq. (\ref{eq:k_disc_01}) simply shows the discretization for the
standing waves of $r=1$ and $r=-1$.
The discrete energy levels have generally twofold degeneracy reflecting
the spin degrees of freedom.
When we consider the asymmetric velocities,~\cite{PhysRevB.85.165430}
vernier-scale-like discrete energy levels are obtained as shown in
Fig. \ref{fig:0704_diH_50nm}.
Note that the integer $L$, then the offset $\delta k$ shows nearly
three-fold oscillations when the nanotube length $N_s$ changes because
$L \simeq [ (N_s + 1) / 3 ]$ for $k_0 \simeq 2 \pi / 3$.
Energy levels for finite-length of $N_s$ and $N_s + 3$ are almost
identical while that of $N_s + 1$ and $N_s + 2$ are generally
different one another, which is confirm in our numerical calculation
(not shown).
Because the analysis above relies on the low energy condition,
deviation such as the intravalley coupling in the same parity state
as shown in Fig. \ref{fig:0704_diH_50nm} could occur for a larger
energy region.

The second type of the boundary is depicted in
Fig. \ref{fig:1D_boundary_0704} (b).
Each A atom in $|t_2| - t_1 + 1 \le \ell \le t_1$ connects to two B
atoms in the body, and each B atom in $1 \le \ell \le |t_2|$ connects
to two A atoms in the body.
This boundary is the minimal boundary in Fig. \ref{fig:0704_50nm}.
The following conditions are imposed for the wavefunctions,
\begin{align}
  &  \phi_{{\rm A} \ell} = 0 ~~\text{at}~~ \ell = - |t_2| + 1, \cdots, |t_2| - t_1. \label{eq:BC_02_A} \\
  &  \phi_{{\rm B} \ell} = 0 ~~\text{at}~~ \ell = - m/d + 1, \cdots, 0. \label{eq:BC_02_B} 
\end{align}
The number of conditions for $\phi_{{\rm A} \ell}$ is $n/d$, and that
for $\phi_{{\rm B} \ell}$ is $m/d$.~\cite{Akhmerov-2008-02}
Because the number of boundary conditions for $\phi_{{\rm A} \ell}$ is
larger than or equal to the number of relevant modes of A-sublattice
at the left end, $n/d \ge |t_2| + 1$, the standing wave for
A-sublattice should be zero at the left end, which corresponds to
``fixed boundary condition'' for the standing waves.
Then we get the condition $c_{-1} = c_{+1}$ so that the envelope
function of the A-sublattice becomes zero at the left end.
This condition, $c_{-1} = c_{+1}$, reflects that only the intravalley
scattering occurs at the ends.
We also have that the phase difference $\theta_{+1} - \theta_{-1}$ is
fixed to be $\pi / 2$ in the linear dispersion region in which the
relation $\Phi(k_0-k) = \Phi(k_0+k) + \pi$ holds.
Therefore, the wavefunction of A-sublattice vanishes
at the left end because of the envelope function $\sin(k \ell)$.
For this case, Eq. (\ref{eq:from_parity_03}) should be satisfied for
both $r=1$ and $r=-1$.
We have the following discretized wavenumbers,
\begin{equation}
  \frac{ k_{+1} - k_{-1} }{2} = \frac{ l \pi + \theta_{+1} - \theta_{-1} }{N_s + 1}, \label{eq:k_disc_02_00}
\end{equation}
for both parity states $p = \pm 1$, where $l$ is an integer.
Note that the left hand side of Eq. (\ref{eq:k_disc_02_00}) is $k$ if
$k = k_{+1} = - k_{-1}$ holds for the case of symmetric Dirac cone in
which the energy is given by Eq. (\ref{eq:energy_k_gap}).
The corresponding expression is shown for the zigzag
nanotubes.~\cite{PhysRevB.65.165431,PhysRevB.84.165427}
For a given $l$, the two parity states has the same wavenumber.
Therefore the two states have the same energy.
In larger energy region,
we may have a parity dependent deviation for the phase difference from
$\pi/2$ such as $\theta_{+1} - \theta_{-1} = \pi/2 - \vartheta_p k$,
where $- \vartheta_p k$ represents the deviation from $\pi/2$ and
$\vartheta_p$ is a coefficient of the deviation.
For this case we have
$k = ( l + 1/2 ) \pi / ( N_s + 1 + \vartheta_p)$, then there are the
parity splitting for the degenerate energy levels,
\begin{equation}
  \Delta \varepsilon 
  = \left| \gamma \right| \frac{\sqrt{3} a}{2 a_z} k 
  \frac{ \delta \vartheta }{ N_s }, \label{eq:paritySplitting}
\end{equation}
where $\delta \vartheta = \vartheta_{p=1} - \vartheta_{p-1}$.
The dependence of the inverse of the length on the energy splitting is
consistent with the numerical calculation in
Fig. \ref{fig:DelE_fourfold-L_0704}.

For the armchair nanotubes, the orthogonal and the minimal boundary
conditions are identical.
(Note that $n/d=m/d=t_1=|t_2|=1$ for the armchair nanotubes.)
Both for $\phi_{{\rm A} \ell}$ and $\phi_{{\rm B} \ell}$, the number
of boundary conditions is one and there are two relevant modes.
Therefore the same discussion with the orthogonal boundary condition
is available.

\subsection{Slowly decaying modes}
\label{sec:SlowlyDecayingMode}

Finally we comment on the number of slowly decaying modes in the
calculation.
For a SWNT which is longer than the slowly decaying modes, the modes
appear at the zero energy.
As shown in \S \ref{sec:1Dmodel_modes}, the number of B-like
evanescent modes, including the two slowly decaying modes, at the left
end is $t_1 + 1$ for the metal-2 nanotubes with $\mu=0$ states.
For the orthogonal boundary, the number of boundary conditions for the
B-sublattice at the left end is $t_1$
Therefore, the number of independent evanescent modes is $t_1 + 1 -
t_1 = 1$ for each spin and each end.
There are total 4 independent evanescent modes for both ends
orthogonal boundary.
Since the longest evanescent mode dominates in each eigenstate, 
4 slowly decaying modes appear.

For the minimal boundary, the number of boundary conditions for the
B-sublattice at the left end is $m/d$, then the number of the
independent evanescent modes is $t_1 + 1 - m/d = (n-m)/3d + 1$ for
each spin and each end.
Since the number of slowly decaying modes is two in the independent
evanescent modes, 8 slowly decaying modes appear for both ends minimal
boundary, as shown in Fig. \ref{fig:0704_50nm}.
The remaining independent evanescent modes with shorter decay length
would also appear for the case of $(n-m)/3d + 1 \ge 3$.

For the SWNTs shorter than the decay length of the slowly decaying
modes, the slowly decaying modes appear at finite energies to satisfy
the boundary conditions, as shown in Figs. \ref{fig:0704_diH_50nm} and
\ref{fig:0704_50nm}.
Note that the A-like evanescent modes do not appear at the left end
because the number of A-like evanescent modes, $|t_2| - 1$, is smaller
than that of the boundary conditions for the above boundaries, $|t_2|$
or $n/d$.

\section{Conclusion}
\label{sec:Conclusion}

In summary, we studied the discrete energy levels in the finite-length
m-SWNTs.
For the metal-1 nanotubes with the $C_d$ rotational symmetry, the two
valleys are decoupled in the eigenfunctions because they have
different orbital angular momenta.
The energy levels have nearly fourfold degeneracy and the spin-orbit
interaction lifts the degeneracy in the order of meV.
For the metal-2 nanotubes, on the other hand, the two valley states
have the same orbital angular momentum and they are strongly coupled
for the orthogonal boundary and the cap-termination as well as the
armchair nanotubes.
The energy levels showed the vernier-like spectrum reflecting the
asymmetric velocities and the strong valley coupling.
For the metal-2 nanotubes with minimal boundary, nearly fourfold
behavior was observed, reflecting nearly decoupling of two valleys.
For this case, the parity splitting overcomes the spin-orbit splitting
for the short nanotubes.
The effective one-dimensional model explained the coupling of the two
valleys, appearance of the slowly decaying modes caused by the
curvature-induced shift of the Dirac point, the length dependence of
the parity splitting.
The spectrum types for nanotube types discussed in this paper are
summarized in Table \ref{table:degeneracy}.

\begin{table}[tbhp]
  \caption{Summary of the degeneracy for the finite-length m-SWNTs
    with $C_d$ rotational symmetry around the tube axis, and with
    $C_2$ rotational symmetry around the axis perpendicular to the
    tube axis.
    Metal-1 nanotubes satisfy the relation $d_R = d$, and metal-2
    nanotubes satisfy the relation $d_R = 3d$.
    OB means the orthogonal boundary, MB means the minimal boundary,
    and SO means spin-orbit.  The armchair nanotubes with armchair
    edges are categorized in Metal-2 (OB).
    Chirality dependence of the spin-orbit splitting for each valley
    is given in Ref. ~\onlinecite{izumida-2009-06}.  }
  \label{table:degeneracy}
  \begin{tabular}{ll}
    \hline \hline
    Type    & Spectrum \\
    \hline
    Metal-1 & nearly fourfold, SO splitting \\
    Metal-2 (OB) & vernier-scale-like \\
    Metal-2 (MB) & nearly fourfold, SO splitting, parity splitting \\
    \hline \hline
  \end{tabular}
\end{table}

In this paper we restrict the finite-length SWNTs without any external
field.
The finite-length effects studied in this paper will directly checked
in devices made of all nano-carbons including SWNTs.
In the many experimental setup with the metal-gated SWNTs, additional
potentials could be created to confine the electrons inside the tubes.
However, the nanotube ends may still influence to the electrons in a
center region because of the Klein tunneling through the
barriers~\cite{Steele-2009} or for the nanotubes on the metal
electrodes.~\cite{Nygard-1999-09}
Furthermore we did not consider effects of Coulomb interaction between
electrons, which can be the same order of the level spacing of the
single-particle energy.
One of the main feature could be captured within the so-called
constant interaction model,~\cite{Sapmaz-2005-04} which simply
increase the label-independent constant term in the addition energy.
The degeneracy behavior studied in this paper could be useful for
understanding the orbital-related correlated electrons such as the
Kondo
effect~\cite{Liang-2002-03,Jarillo-Herrero-2005-03,Makarovski-2007-08,PhysRevLett.111.136803}
and the Pauli blockade.~\cite{Pei:2012fk}
The interaction could also cause intervalley scattering as well as
intravalley scattering, and might affects especially on the valley
decoupling features.
The effective 1D model derived in this paper could be a lattice model
to treat the Coulomb interaction for the given geometry of the
chirality with the boundary for the recent observed 1D correlated
electron effects such as the Mott insulator~\cite{Deshpande02012009}
and the Wigner crystallization.~\cite{Deshpande-2008-04,Pecker:2013kx}
The effects of additional potential effects as well as the Coulomb
interaction in the finite-length SWNTs with boundaries should be
clarified in future study.

\begin{acknowledgments}
  We acknowledge MEXT Grants (Nos. 22740191, 26400307, 15K05118 for
  W.I, Nos. 25107001, 25107005 for R.O. and R.S., No. 25286005 for
  R.S.), Japan.
  We would like to thank to Y. Tatsumi and M. Mizuno for help on
  computational calculation.
\end{acknowledgments}

\appendix

\section{Long cutting lines passing through $K$ and $K'$ points for metal-1 and metal-2 nanotubes}
\label{sec:App:LongCuttingLine}

Here, we will show the fact that a long cutting line passes through
both $K$ and $K'$ points for the metal-2 nanotubes, whereas cutting
line passes only through $K$ or $K'$ points for the metal-1 nanotubes.

If a cutting line passes through both $K$ and $K'$ points, the cutting
line should also passes through a $\Gamma$ point because the $K'$
point sits on the opposite side of $K$ point with respect to the
$\Gamma$ point.
Therefore, the cutting line should be $\mu=0$.
If the $\mu=0$ cutting line does not pass a $K$ point, there is no
cutting line which passes both $K$ and $K'$ points for a given
$(n,m)$.

The condition that the $\mu=0$ cutting line passes through a $K$ point
is expressed by,
\begin{equation}
  \left( \alpha + \frac{1}{3} \beta \right) \bm{K}_2
  = \overrightarrow{\Gamma K} + j_1 \bm{b}_1 + j_2 \bm{b}_2,
  \label{eq:app:KpassingCond_00}
\end{equation}
where $\overrightarrow{\Gamma K} = (2 \bm{b}_1 + \bm{b}_2)/3$ is the
vector from $\Gamma$ point to $K$ point in a hexagonal BZ, $\alpha$,
$\beta$, $j_1$ and $j_2$ are integers.
Here $\beta$ is introduced as follows; 
\begin{equation}
  \beta = \left\{
  \begin{array}{cl}
    0  & \text{~~for metal-1}, \\
    +1 & \text{~~for metal-2$p$}, \\
    -1 & \text{~~for metal-2$m$}. \\
  \end{array}
  \right.
  \label{eq:app:beta_definition}
\end{equation}
In Eq. (\ref{eq:app:KpassingCond_00}), we used the already known fact
that a $K$ point is mapped onto the center of a short cutting line for
the metal-1 nanotubes, whereas it is mapped onto 5/6 (1/6) position of
a short cutting line for the metal-2$p$ metal-2$m$
nanotubes.~\cite{Saito-2005-10}
It should be noted that the following relation holds for the metal-2
nanotubes,
\begin{equation}
  {\rm mod} \left( \frac{n}{d}, 3 \right) = {\rm mod} \left( \frac{m}{d}, 3 \right),
  \label{eq:app:metal2Cond_02}
\end{equation}
because of the relation $(n-m)/3d = [(2n+m) - (2m+n)] / d_R = -t_1 -
t_2$.
In order to satisfy Eq. (\ref{eq:app:KpassingCond_00}), $j_1$, $j_2$ and
$\alpha$ should satisfy the following equations.
By comparing the coefficients of $\bm{b}_1$ and $\bm{b}_2$ in
Eq. (\ref{eq:app:KpassingCond_00}), we have,
\begin{align}
  & j_1 n + j_2 m = - \frac{1}{3} \left( 2 n + m \right), \label{eq:app:KpassingCond_01} \\
  & \alpha = \left( \frac{2}{3} + j_1 \right) \frac{N}{m} - \frac{\beta}{3}. \label{eq:app:KpassingCond_02}
\end{align}

For the case of metal-2 nanotubes, it will be shown that the following
$j_1$ and $j_2$ satisfy the conditions Eqs. (\ref{eq:app:KpassingCond_01})
and (\ref{eq:app:KpassingCond_02}),
\begin{align}
  & j_1 = - \frac{1}{3} \left( \beta \frac{m}{d} + 2 \right), \label{eq:app:j_1} \\
  & j_2 = - \frac{1}{3} \left( - \beta \frac{n}{d} + 1 \right). \label{eq:app:j_2}
\end{align}
Note that the right hand sides of Eqs. (\ref{eq:app:j_1}) and
(\ref{eq:app:j_2}) are integers because of Eqs. (\ref{eq:metal2pmCond_01})
and (\ref{eq:app:metal2Cond_02}) for the metal-2 nanotubes.
By substituting Eq. (\ref{eq:app:j_1}) for Eq. (\ref{eq:app:KpassingCond_02}),
$\alpha$ is given by
\begin{equation}
  \alpha = - \frac{\beta}{3} \left( \frac{N}{d} + 1 \right). \label{eq:app:alpha_02}
\end{equation}
Because $N = 2 (n^2 + m^2 + nm) / d_R = [(2m + n)(2n + m) - 3 nm] /
d_R$ and $d_R = 3 d$ for the metal-2 nanotubes, the following relation
folds,
\begin{equation}
  \frac{N}{d} = - 3 t_1 t_2 - \frac{n m}{d^2}.
\end{equation}
Using this, Eq. (\ref{eq:app:alpha_02}) becomes
\begin{equation}
  \alpha = \beta t_1 t_2 + \frac{\beta}{3} \left( \frac{n m}{d^2} - 1 \right). \label{eq:app:alpha_03}
\end{equation}
The right hand side of Eq. (\ref{eq:app:alpha_03}) is an integer
because of the relation ${\rm mod} ( nm/d^2, 3 ) = 1$, which can be
derived from Eqs. (\ref{eq:metal2pmCond_01}) and
(\ref{eq:app:metal2Cond_02}).
Therefore, Eq. (\ref{eq:app:KpassingCond_00}) is satisfied by the set of
integers of Eqs. (\ref{eq:app:j_1}), (\ref{eq:app:j_2}) and
(\ref{eq:app:alpha_03}) for the metal-2 nanotubes.
From the left hand side of Eq. (\ref{eq:app:KpassingCond_00}) and
Eq. (\ref{eq:app:alpha_02}), it is shown that the $K$ point is located at
$1/6$ ($5/6$) position of the longer cutting line defined by
Eqs. (\ref{eq:cuttingLine}) and (\ref{eq:cuttingLine_region_long})
with $\mu=0$ for the metal-2$p$ (metal-2$m$) nanotubes.

For the case of metal-1 nanotubes, dividing Eq. (\ref{eq:app:KpassingCond_01})
by $d$, one gets,
\begin{equation}
  j_1 \frac{n}{d} + j_2 \frac{m}{d} = \frac{1}{3} t_2. \label{eq:app:KpassingCond_metal1_01}
\end{equation}
Because $t_2$ is not a multiple of 3 for the metal-1 nanotubes,
any integers of $j_1$ and $j_2$ cannot satisfy
Eq. (\ref{eq:app:KpassingCond_metal1_01}).
Therefore, there is no cutting line passing through both $K$ and $K'$
points for the metal-1 nanotubes.

\section{Numerical calculation for metal-1, armchair and capped metal-2 nanotubes}
\label{sec:App:numCalc}

Here we will show some numerical calculation of the finite-length
metal-1, armchair, and capped metal-2 nanotubes using the extended
tight-binding model.
As expected, the metal-1 with $C_d$ rotational symmetry will exhibit
nearly fourfold degeneracy and its lift by the spin-orbit interaction.
On the other hand, the armchair and the capped metal-2 nanotubes will
exhibit the vernier-like spectra as well as
Fig. \ref{fig:0704_diH_50nm}.

\subsection{Energy levels for metal-1 nanotubes}
\label{sec:app:numCalMetal1}

\begin{figure}[htb]
  \includegraphics[width=8cm]{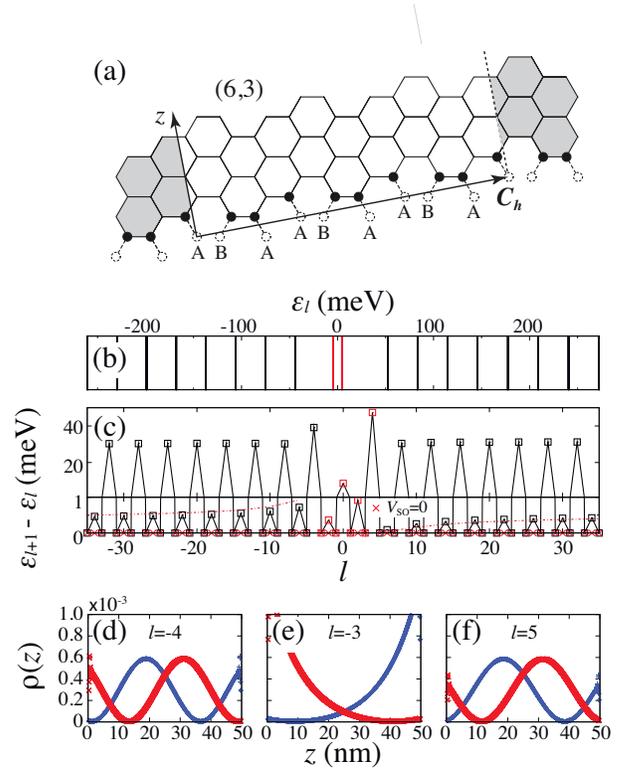}
  \caption{(Color online) Boundary shape, calculated energy levels and
    eigenstates for $(6,3)$ nanotube of $50.15$ nm length with 
    both ends minimal boundary as shown in (a).
    (b) Energy levels $\varepsilon_l$ in $-35 \le l \le 35$.
    (c) Level separation, $\varepsilon_{l+1} - \varepsilon_l$, as a
    function of $l$.  
    The dashed lines in the lower panel show the spin-orbit splitting
    for corresponding energy calculated by the energy band
    calculation.
    (d)-(f) Local density for (c) $l=-4$, (d) $l=-3$ and (e)
    $l=5$.  
  }
  \label{fig:0603_50nm}
\end{figure}

Figure \ref{fig:0603_50nm} (b) shows the energy levels for (6,3)
nanotube with $50.15$ nm length with the minimal boundary for both
ends keeping $C_3$ ($d=3$) rotational symmetry.
The left end is depicted in Fig. \ref{fig:0603_50nm} (a).
There are the slowly decaying modes ($-3 \le l \le 4$) in the energy
gap between $\varepsilon_{l=-4} = -44$ meV and $\varepsilon_{l=5} =
53$ meV.
Above and below the energy gap, the level separation shows almost
equal interval reflecting the quantization of the linear energy
dispersion.
Fig. \ref{fig:0603_50nm} (c) shows that each levels show nearly
fourfold degeneracy.
The levels show complete fourfold degeneracy for the case of absence
of spin-orbit interaction.
The spin-orbit interaction lifts the fourfold degeneracy as expected
in the energy band calculation.~\cite{izumida-2009-06}
Other metal-1 nanotubes, for instance, metallic (9,0)-zigzag
nanotubes, also show similar behaviors with Fig. \ref{fig:0603_50nm},
nearly fourfold degeneracy and lift of the degeneracy by the
spin-orbit interaction (not shown).
In addition, finite length (9,0) nanotubes show the edge states
decaying in the length scale of carbon-carbon bond, not on but below
the charge neutral point, as the effect of the next nearest-neighbor
hopping process.~\cite{Sasaki-2006-11}

\begin{figure}[htb]
  \includegraphics[width=8cm]{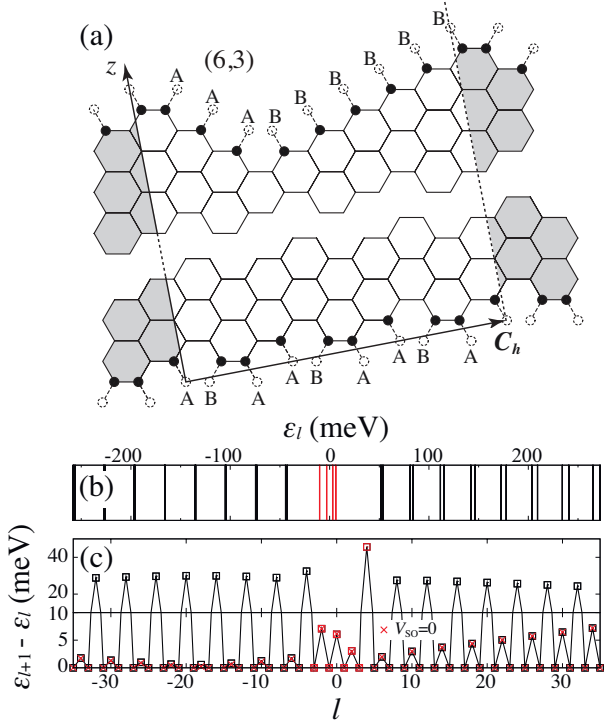}
  \caption{(Color online) Boundary shape and calculated energy levels
    $\varepsilon_l$ for $(6,3)$ nanotube of $50.63$ nm length.
    (a) Unfolded tube near the left (lower) and right (upper) ends.
    The upper end is cut along the $n \bm{a}_1$ and $m \bm{a}_2$.
    Note that both ends are classified into the minimal boundary.
    (b) Energy levels $\varepsilon_l$ in $-35 \le l \le 35$.
    (c) Level separation, $\varepsilon_{l+1} - \varepsilon_l$, as a
    function of $l$.
  }
  \label{fig:0603_pcut_nmcut_50nm}
\end{figure}

When the rotational symmetry at the end is broken, the valley
degeneracy and its lift by the spin-orbit interaction will not be
clearly observed.
Figure \ref{fig:0603_pcut_nmcut_50nm} shows the energy level for (6,3)
nanotube with $50.63$ nm length, in which the one side of the end is
cut along the $n \bm{a}_1$ and $m \bm{a}_2$ and the other side is the
same with Fig. \ref{fig:0603_50nm}.
The system loses the $C_3$ rotational symmetry which is possessed in
the case of Fig. \ref{fig:0603_50nm}.
The large lift of the fourfold degeneracy, for instance, $\Delta
\epsilon \gsim$ 6 meV in $\varepsilon \gsim 200$ meV, is not caused by
the spin-orbit interaction, but due to the mixing of the two valley
degrees of freedom by breaking of the rotational symmetry.

In the end of this subsection, we comment on valley mixing effect by
the spin-orbit interaction.
Strictly speaking, the spin-orbit interaction could mix the states in
two valleys because spin-up states in $\mu=1$ cutting line, which
passes the $K'$ point, and spin-down states in $\mu=2$ cutting line,
which passes the $K$ point, have the same total angular momentum,
$3/2$ (see Fig. \ref{fig:cuttingLine} (a) for the cutting lines).
Mixing of the two valleys may give an additional effect from the
spin-orbit splitting, such as the parity splitting in
Fig. \ref{fig:0704_50nm} or the splitting in
Fig. \ref{fig:0603_pcut_nmcut_50nm} for the absence of $C_3$
rotational symmetry.
Such valley mixing effects, however, seem to be irrelevant in
Fig. \ref{fig:0603_50nm}, which simply shows the splitting as expected
from the band calculation.

\subsection{Vernier spectra for (6,6)-armchair and capped (7,4) nanotubes}
\label{sec:app:vernier}

\begin{figure}[htb]
  \includegraphics[width=8cm]{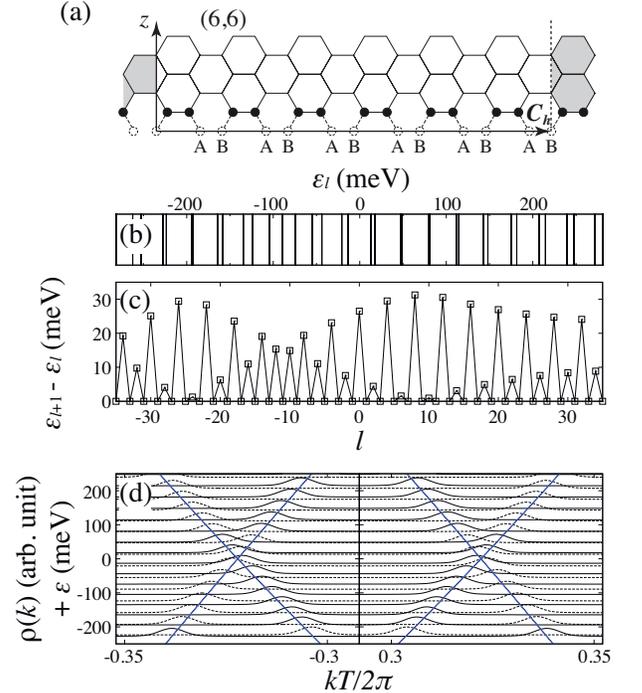}
  \caption{(Color online) Boundary shape, calculated energy levels and
    intensity plot in wavenumber for $(6,6)$ armchair nanotube of
    $50.05$ nm length.
    (a) Unfolded tube near the left end.
    (b) Energy levels in $\varepsilon_{l=-35} \le \varepsilon \le
    \varepsilon_{l=35}$.
    (b) Level separation, $\varepsilon_{l+1} - \varepsilon_l$, as a
    function of $l$.
    (c) Intensity plot of Fourier transform of wavefunction on
    A-sublattice for each level.  The intensities for the states of
    spin-up-majority are shown.  The blue lines show the energy band
    calculation under the periodic boundary condition.
  }
  \label{fig:0606_50nm}
\end{figure}

Figure \ref{fig:0606_50nm} shows the energy levels for (6,6)-armchair
nanotube with $50.05$ nm length.
The energy levels show the similar behaviors with the vernier-like
spectrum in Fig. \ref{fig:0704_diH_50nm}.
This can be understood by (i) the strong intervalley coupling and
(ii) the asymmetric velocities, as well as
Fig. \ref{fig:0704_diH_50nm}.
For the armchair nanotubes, it can be shown using the effective 1D
model that even channel from $(a_\ell + b_\ell)/\sqrt{2}$ states and
odd channel from $(a_\ell - b_\ell)/\sqrt{2}$ states are decoupled
each other for this boundary, and the even (odd) channel has the
energy band with left-going states at the $K$-valley ($K'$-valley) and
right-going states at the $K'$-valley ($K$-valley).
Therefore, no intravalley mixing is seen in Fig. \ref{fig:0606_50nm}
(c).
The period of the two- to fourfold oscillation is not constant but has
the energy dependence, for instance, the period becomes longer for the
positive energy region.
This is because the velocities has energy dependence reflecting the
deviation from the linear energy band.
The velocity difference between left- and right-going waves becomes
smaller for the higher energy region in the conduction band
as well as the (7,4) nanotube.

\begin{figure}[htb]
  \includegraphics[width=8cm]{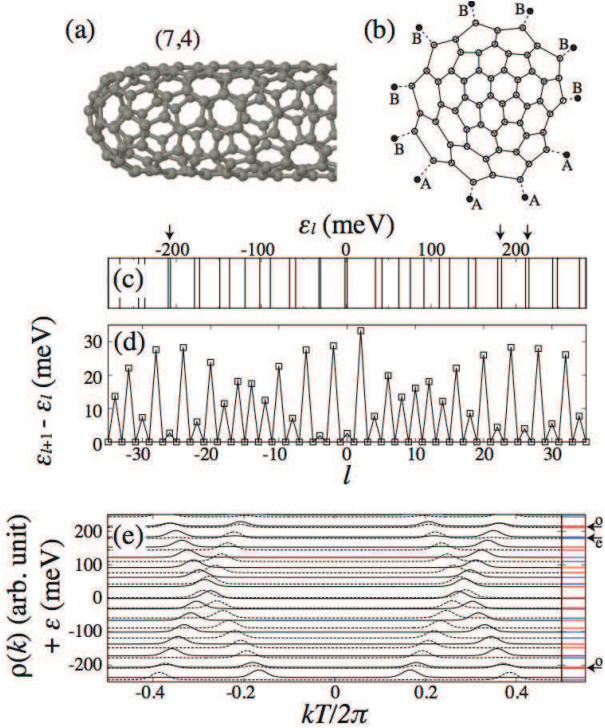}
  \caption{(Color online) Boundary shape, calculated energy levels and
    intensity plot in wavenumber for $(7,4)$ nanotube of $49.56$ nm
    length with both-ends-capped.
    (a) 3D and (b) 2D representation of the cap structure.  In (b),
    the solid circles indicate the A and B atoms connected to the cap
    region, and the open circle indicate the carbon atoms in the cap
    region at the left end.
    (c) Energy levels $\varepsilon_l$ in $-35 \le l \le 35$.
    (d) Level separation $\varepsilon_{l+1} - \varepsilon_l$ as a
    function of $l$.  
    (e) Intensity plot of Fourier transform of wavefunction on
    A-sublattice for each level.
    Right figure in (e) shows the energy levels of the even parity
    (blue lines) and the odd parity (red lines) for $V_{\rm SO}=0$.
    The arrows in (c) and (e) show the states exhibiting intravalley
    coupling.}
  \label{fig:0704_cap_50nm}
\end{figure}

We show another case exhibiting the vernier-like spectrum for a capped
nanotube, which would be more abundant than the orthogonal boundary
containing the Klein-terminations.
Figure \ref{fig:0704_cap_50nm} shows the calculated energy levels for
both-side-capped (7,4) nanotubes of $49.56$ nm length.
In the calculation, the cap structure is formed by using the graph
theory,~\cite{Brinkmann200255,Brinkmann-CaGe-2010} in which the cap
region is defined outside the end cut along the $n \bm{a}_1$ and $m
\bm{a}_2$.
There are two cap obeying the isolated pentagon rule for (7,4)
nanotube, one has 55 and the other has 57 carbon atoms at the cap
region.
The cap with 57 carbon atoms is used in the calculation.
To connect the cap and the body smoothly, the structure is optimized
by the molecular mechanics method with the Universal force field (UFF)
in the Gaussian program.~\cite{g09}
Even the optimization method gives less accuracy on the electronic
fine structure such as the curvature-induced energy gap and the
spin-orbit interaction, we could discuss the intervalley coupling in
the eigenfunctions from the semi-quantitative calculation.
The vernier-like spectrum is seen in the calculated energy levels.
As shown in the plot of Fourier transform, in general, each level is
formed from a left-going wave of one valley and a right-going wave of
another valley, that corresponds to $|c_r/c_{-r}| \ll 1$ ($r=1$ or
$-1$) in Eqs. (\ref{eq:standingWave_A}) and (\ref{eq:standingWave_B})
reflecting the strong intervalley coupling.
For closer two levels, the intravalley mixing between the same parity
states is also seen, for instance, $\varepsilon_l \sim -208$ meV,
$181$ and $213$ meV.
The vernier-like spectrum similar with Fig. \ref{fig:0704_cap_50nm} is
also obtained for another cap obeying the isolated pentagon rule with
55 carbon atoms in the numerical calculation (not shown).

The strong valley coupling for the case of cap-termination may be
understood in the 1D model as follows.
The boundary conditions at the left end can be given by $\phi_{{\rm A}
  \ell} = \phi_{{\rm c} \ell_{\rm c}}$ for $n$ sets of $(\ell_{\rm
  A},\ell_{\rm c})$ and $\phi_{{\rm B} \ell} = \phi_{{\rm c} \ell_{\rm
    c}}$ for $m$ sets of $(\ell_{\rm A},\ell_{\rm c})$, where
$\ell_{\rm A}$ ($\ell_{\rm B}$) is the index of the coordinate of
A-sublattice (B-sublattice)
and $\ell_{\rm c}$ is that of cap.
For the amplitudes $\phi_{{\rm c} \ell_{\rm c}}$, there are $N_{\rm
  c}$ equations of motion.
In general, $\phi_{{\rm c} \ell_{\rm c}}$ shows oscillation in the
length scale of carbon-carbon bond.
Therefore, one also expects the fast oscillations for the
wavefunctions of A- and B-sublattices, which can be formed under the
strong intervalley coupling of $|c_r/c_{-r}| \ll 1$ for $r=1$ or $-1$.

\section{Derivation of effective 1D Model}
\label{sec:App:Derivation1Dmodel}

Here we will show the detailed of the derivation of the effective 1D
model given in \S \ref{sec:1DmodelDerivation}.

\begin{figure}[htb]
  \includegraphics[width=8cm]{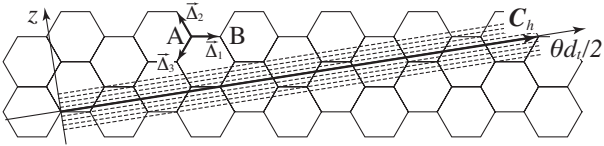}
  \caption{Coordinates for $(7,4)$ nanotube.  The dashed lines show
    the interval of the 1D lattice constant $a_z = T d / N$.  For this
    case, $d=1$, $T=\sqrt{31}a$ and $N=62$ .
  }
  \label{fig:Coordinates_0704}
\end{figure}

In the cylindrical coordinate system in which the tube axis coincides
the $z$ axis, the components of $\vec{\Delta}_j$ appearing in
Eq. (\ref{eq:H_00}) ($\vec{\Delta}_j$ is depicted in
Fig. \ref{fig:Coordinates_0704}) is given by
\begin{align}
  & \Delta\theta_1 =   \pi \frac{n + m}{n^2 + m^2 + nm}, \\
  & \Delta\theta_2 = - \pi \frac{m}{n^2 + m^2 + nm},     \\
  & \Delta\theta_3 = - \pi \frac{n}{n^2 + m^2 + nm},
\end{align}
and
\begin{align}
  & \Delta z_1 =  - a_z \frac{n - m}{3d},  \label{eq:app:Deltaz1_00} \\
  & \Delta z_2 =    a_z \frac{2n + m}{3d}, \label{eq:app:Deltaz2_00} \\
  & \Delta z_3 =  - a_z \frac{2m + n}{3d}. \label{eq:app:Deltaz3_00}
\end{align}

Let us consider the following Fourier transform in the circumference
direction for the operators,
\begin{equation}
  c_{\sigma (\mu, z)}
  = \frac{1}{\sqrt{d}} 
  \sum_{\vec{r}} {\rm e}^{-{\rm i} \mu \theta_{\vec{r}}} c_{\sigma {\vec{r}}} \delta_{z,z_{\vec{r}}},
\end{equation}
where $\sigma = {\rm A, B}$, and $z = a_z \ell$ denotes the lattice
position of A atoms on the nanotube axis, $\ell$ is the integer for
the lattice position.
The summation on $\vec{r}$ is taken place for a given $z$.
The inverse Fourier transform is given by
\begin{equation}
  c_{\sigma {\vec{r}}} 
  = \frac{1}{\sqrt{d}} 
  \sum_{\mu=0}^{d-1} {\rm e}^{{\rm i} \mu \theta_{\vec{r}}} c_{\sigma (\mu, z_{\vec{r}})}. \label{eq:app:invFourier_C}
\end{equation}
Substituting Eq. (\ref{eq:app:invFourier_C}) for the Hamiltonian
Eq. (\ref{eq:H_00}), the Hamiltonian can be decomposed into projected
Hamiltonian for $\mu$-th angular momentum $H_\mu$,
\begin{align}
  & H = \sum_\mu H_\mu, \\
  & H_\mu = 
  \sum_z \sum_{j=1}^3 \gamma_j {\rm e}^{{\rm i} \mu \Delta\theta_j}
  c_{{\rm A} (\mu, z)}^\dagger  c_{{\rm B} (\mu, z + \Delta z_j)} + {\rm H.c.},
\end{align}
By selecting the $\mu$ values in which $H_\mu$ contains states near
the Fermi energy, one gets an effective 1D Hamiltonian.

Because of our main interest, 
let us focus on the metal-2 nanotubes.
For the metal-2 nanotubes, $\Delta z_j$ can be expressed by
\begin{align}
  & \Delta z_1 =    a_z (t_1 + t_2), \label{eq:app:Deltaz1} \\
  & \Delta z_2 =  - a_z t_2,         \label{eq:app:Deltaz2} \\
  & \Delta z_3 =  - a_z t_1.         \label{eq:app:Deltaz3}
\end{align}
It should be mentioned that there is a B atom at $\beta \bm{C}_h /
d_R$ from each A atom, that is, there are $d$ pairs of A and B atoms
on the same $z$ for the metal-2 nanotubes.
[For example, see Fig. \ref{fig:Coordinates_0704} for the coordinates
  of $(n,m)=(7,4)$ metal-2$p$ nanotube ($d=1$).]
We consider only $H_{\mu=0}$ and use the simplified notation,
\begin{equation}
  c_{{\rm A} (\mu=0,z=a_z \ell)} \rightarrow a_{\ell}, ~~~~
  c_{{\rm B} (\mu=0,z=a_z \ell)} \rightarrow b_{\ell}.
\end{equation}
and $\Delta \ell_j = \Delta z_j / a_z$, then we get
Eq. (\ref{eq:H1D_metal2}) for the effective 1D Hamiltonian for the
metal-2 nanotubes.

\section{Mode analysis of effective 1D model}
\label{sec:App:1Dmodel_modes}

Here we will discuss the detail derivation of the modes of
Eqs. (\ref{eq:EOM_A_alg}) and (\ref{eq:EOM_B_alg}).

As a general property, when $(\lambda, \eta)$ is a set of solution of
Eqs. (\ref{eq:EOM_A_alg}) and (\ref{eq:EOM_B_alg}), $(1/\lambda,
1/\eta)$ is another set of solution since the equations are equivalent
by changing $(\lambda, \eta) \leftrightarrow (1/\lambda, 1/\eta)$.
For a solution $(\lambda, \eta)$, the complex conjugates, $(\lambda^*,
\eta^*)$ is also another set of solution since $\gamma_j$ and
$\varepsilon$ are real numbers.

We will start from the traveling modes ($|\lambda| = |\eta| = 1$) in
the solutions of Eqs. (\ref{eq:EOM_A_alg}) and (\ref{eq:EOM_B_alg}).
For the traveling modes, Eqs. (\ref{eq:EOM_A_alg}) and
(\ref{eq:EOM_B_alg}) are equivalent as mutual complex conjugate.
Let us first consider the flat graphene case, $\delta\gamma_j = 0$.
By noting the fact that the left hand side of Eq. (\ref{eq:EOM_A_alg})
is zero for $\lambda = e^{\pm {\rm i} 2 \pi / 3}$ because of the
relations Eqs. (\ref{eq:metal2pmCond_01}) and
(\ref{eq:app:metal2Cond_02}), we can expand at the wavenumbers $k_\ell
= \pm 2 \pi /3$ which correspond to the two Dirac points.
It is shown that there are four traveling modes for a given energy;
$(\lambda, \eta) = \left( e^{{\rm i} k_\ell}, e^{{\rm i} \Phi(k_\ell)}
\right)$, where $k_\ell = \tau' 2 \pi / 3 + k'$
denotes four wavenumbers for a given energy.
We restricted in the linear dispersion regime, in which the energy and
$k'$ has the relation, 
\begin{equation}
  \varepsilon = \pm |\gamma| \frac{ \sqrt{3} a}{2 a_z} |k'|.
\end{equation}    
The phase in Eq. (\ref{eq:travelingModes_gap}) is given by
\begin{equation}
  \Phi \left( \tau' \frac{2 \pi}{3} + k' \right) 
  = \tau' 
  \left( 
    \frac{2 \pi}{3} t_2 + \beta \theta 
    \right) 
    + \arg \left( \frac{\gamma k'}{{\rm i} \varepsilon} \right),
\end{equation}    
$\beta = \pm 1$ is the value defined in
Eq. (\ref{eq:app:beta_definition}) for the metal-2 nanotubes, and
$\theta$ is the chiral angle.
Because of the time-reversal symmetry, we have $\Phi(k_\ell) = -
\Phi(- k_\ell)$.

When $\delta\gamma_j \ne 0$, the above analysis should be modified as
follows.
For a state with wavenumber $k_\ell = \tau' 2 \pi / 3 + k'$, by
considering the contribution of lowest order of $k'$ and
$\delta\gamma_j / \gamma$, Eq. (\ref{eq:EOM_A_alg}) is written as,
\begin{equation}
  - {\rm i} \gamma e^{\tau' {\rm i} \left( \frac{2 \pi}{3} t_2 + \beta \theta \right) }
  \frac{\sqrt{3} a}{2 a_z}
  \left[ \left( k' - \tau' k_{\rm R} \right) + {\rm i} k_{\rm I} \right]
  = \varepsilon \eta.
  \label{eq:App:EOM_A_alg_traveling_01}
\end{equation}
Here $k_{\rm R}$ and $k_{\rm I}$ satisfy the following relations,
\begin{align}
  k_{\rm R} 
  & = 
  \frac{2 a_z}{\sqrt{3} a} \beta \sum_{j=1}^3 \frac{\delta\gamma_j}{\gamma} 
  \sin \left[ \theta + \frac{2 \pi}{3} \left( j-2 \right) \right], \\
  k_{\rm I} 
  & = 
  \frac{2 a_z}{\sqrt{3} a} \sum_{j=1}^3 \frac{\delta\gamma_j}{\gamma} 
  \cos \left[ \theta + \frac{2 \pi}{3} \left( j-2 \right) \right].
\end{align}
$k_{\rm R}$ and $k_{\rm I}$ relate to the shift of the Dirac point in
$\bm{K}_1$ and $\bm{K}_2$ direction, respectively, from the $K$ or
$K'$ points in 2D BZ because of the curvature of nanotube surface.
By comparing with the energy band calculation with the extended
tight-binding method,~\cite{izumida-2009-06} they have the relations
in Eq. (\ref{eq:kR_kI_compare_bandCalc}).
From Eq. (\ref{eq:App:EOM_A_alg_traveling_01}), it is shown that there
are the following four traveling modes,
\begin{equation}
  (\lambda, \eta) = \left( e^{{\rm i} k_\ell}, e^{{\rm i} \Phi(k_\ell)} \right), ~~~~
  k_\ell = \tau' k_0 + k, \label{eq:App:travelingModes_gap}
\end{equation}
for the energy outside the energy gap, $|\varepsilon| > \varepsilon_{\rm gap}$,
where the energy gap 
\begin{equation}
  \varepsilon_{\rm gap} 
  =
  |\gamma| \frac{\sqrt{3} a}{2 a_z} |k_{\rm I}|
\end{equation}
is induced by the curvature of nanotube
surface.~\cite{Kane-1997-03,Ando-2000-06,izumida-2009-06}
After Eq. (\ref{eq:App:travelingModes_gap}), $k$ is the wavenumber
measured from $\tau' k_0 = \tau' \left( 2 \pi / 3 + k_{\rm R}
\right)$, which is the bottom (top) position of the conduction
(valence) band.
The energy and $k$ has the following relation
\begin{equation}
  \varepsilon = \pm |\gamma| \frac{\sqrt{3} a}{2 a_z} \sqrt{ k^2 + k_{\rm I}^2 },
  \label{eq:App:energy_k_gap}
\end{equation}
and the phase is given by
\begin{equation}
  \Phi(\tau' k_0 + k) 
  = \tau' \left( \frac{2 \pi}{3} t_2 + \beta \theta \right)
  + \arg \left[ \frac{\gamma \left( k + {\rm i} k_{\rm I} \right) }{{\rm i} \varepsilon} \right].
  \label{eq:App:phase_gap}
\end{equation}

Inside the gap, $|\varepsilon| < \varepsilon_{\rm gap}$, there are no
traveling modes, but four slowly decaying evanescent modes
exist.
Near the band edges, the modes can be analyzed by changing $\left( k'
- \tau' k_{\rm R} \right) \rightarrow i \kappa$ in
Eq. (\ref{eq:App:EOM_A_alg_traveling_01}) and in a pair equation
derived from (\ref{eq:EOM_B_alg}).
Then it is shown straightforwardly that the modes $( e^{i \tau' k_0 -
  \kappa}, \eta )$ have the energy
\begin{equation}
  \varepsilon = \pm |\gamma| \frac{\sqrt{3} a}{2 a_z} \sqrt{ k_{\rm I}^2 -\kappa^2 }. 
  \label{eq:App:energy_InsideGap}
\end{equation}
Note that $\kappa$ can be either positive or negative.
From Eqs. (\ref{eq:App:EOM_A_alg_traveling_01}) and
(\ref{eq:App:energy_InsideGap}), we have
\begin{equation}
  | \eta | = \frac{ | k_{\rm I} + \kappa | }{ \sqrt{ k_{\rm I}^2 - \kappa^2 }  }.
\end{equation}

The remaining $2(n/d + m/d) - 4$ modes can be classified into $(n/d +
m/d - 2)$ A-like modes and $(n/d + m/d - 2)$ B-like modes.
For the A-like modes, using $|\varepsilon / \gamma| \ll 1$,
$|\delta\gamma_j/\gamma| \ll 1$ and $|\eta| < 1$,
Eqs. (\ref{eq:EOM_A_alg}) and (\ref{eq:EOM_B_alg}) can be simplified
as
\begin{align}
  \lambda^{n/d + m/d} + \lambda^{n/d} + 1 = 0, \label{eq:App:EOM_A-like_alg_01} \\
  \eta = \frac{\varepsilon}{\gamma} \frac{1}{\lambda^{- (|t_2| - t_1)} + \lambda^{|t_2|} + \lambda^{-t_1}}. \label{eq:App:EOM_A-like_alg_02}
\end{align}
The first equation (\ref{eq:App:EOM_A-like_alg_01}) has the roots
$\lambda = e^{\pm {\rm i} 2 \pi / 3}$ and the properties for these
modes have already been captured as the traveling modes or the slowly
decaying modes in the previous paragraphs.
For the remaining $(n/d + m/d - 2)$ solutions, the analysis given by
Akhmerov and Beenakker~\cite{Akhmerov-2008-02} is applicable.
Note that Eq. (\ref{eq:App:EOM_A-like_alg_01}) for $\lambda' =
\lambda^{n/d + m/d}$ is the same with Eq. (3.7a) in
Ref. ~\onlinecite{Akhmerov-2008-02}.
By following Appendix B of Ref. ~\onlinecite{Akhmerov-2008-02}, it is
shown that there are $|t_2| - 1$ roots which satisfy $| \lambda | < 1$
(evanescent modes at the left side), and $t_1 - 1$ roots which satisfy
$| \lambda | > 1$ (evanescent modes at the right side) for
Eq. (\ref{eq:App:EOM_A-like_alg_01}).
The second equation (\ref{eq:App:EOM_A-like_alg_02}) determines $\eta$
for each $\lambda$.
The similar discussion can be done for the B-like modes with the
equations which are given by changing $n \leftrightarrow m$ (then $t_1
\leftrightarrow |t_2|$) and $\eta \rightarrow 1/\eta$ in
Eqs. (\ref{eq:App:EOM_A-like_alg_01}) and
(\ref{eq:App:EOM_A-like_alg_02}).
Then it is shown that there are $t_1 - 1$ modes which satisfy $|
\lambda | < 1$, and there are $|t_2| - 1$ modes which satisfy $|
\lambda | > 1$ for the B-like modes.

\section{Traveling modes for A- and B-sublattices for a given boundary}
\label{sec:App:travelingMode}

Here we will show that one of the following relations should be hold
for the orthogonal boundary: $c_{+1}=0$ and $\theta_{-1}$ is
determined, or, $c_{-1}=0$ and $\theta_{+1}$ determined in
Eqs. (\ref{eq:standingWave_A}) and (\ref{eq:standingWave_B}).

Let us first show a useful relation.
The left hand side of Eq. (\ref{eq:EOM_A_alg}) is rewritten as
\begin{align}
  & \frac{1}{\lambda^{|t_2|}}
  \prod_{\tau'=\pm} \left( \lambda - \lambda_{\tau'} \right)
  \prod_{m_1 = 1}^{t_2 - 1} \left( \lambda - \lambda_{m_1}^< \right)
  \prod_{m_2 = 1}^{|t_1| - 1} \left( \lambda - \lambda_{m_2}^> \right), \nonumber \\
  = & 
  \frac{1}{\lambda} 
  \prod_{\tau'=\pm} \left( \lambda - \lambda_{\tau'} \right)
  \prod_{m_1 = 1}^{t_2 - 1} \left( -\lambda_{m_1}^< \right) \nonumber \\
  & \times \prod_{m_1 = 1}^{t_2 - 1} \left( \frac{1}{\lambda} - \frac{1}{\lambda_{m_1}^<} \right) 
  \prod_{m_2 = 1}^{|t_1| - 1} \left( \lambda - \lambda_{m_2}^> \right), \label{eq:app:EOM_A_alg_LHS}
\end{align}
where $\lambda_{\tau'} = e^{{\rm i} \tau' k_0}$, $\lambda_{m_1}^<$,
and $\lambda_{m_2}^>$ are the roots of the left hand side, and we have
$|\lambda_{m_1}^<| < 1$, $|\lambda_{m_2}^>| > 1$.
In Eq. (\ref{eq:app:EOM_A_alg_LHS}), it is shown that the value
$\prod_{m_1 = 1}^{t_2 - 1} \left( -\lambda_{m_1}^< \right)$ is a real
number because there is another root of $\lambda_{m_1}^{<*}$ for a
complex root $\lambda_{m_1}^<$.
For a traveling mode of $\lambda$ which has almost the same
wavenumber with $k_0$ or $- k_0$, $\lambda^{-1} \prod_{\tau'=\pm}
\left( \lambda - \lambda_{\tau'} \right)$ is a small real number.
Let us explicitly consider a traveling mode close to $\lambda_+$,
$(\lambda_+, e^{{\rm i} \Phi})$.
From Eqs. (\ref{eq:EOM_A_alg}) and (\ref{eq:app:EOM_A_alg_LHS}), we
have the following relation, which will be used later,
\begin{equation}
  \prod_{m_1 = 1}^{t_2 - 1} \left( \frac{1}{\lambda_+} - \frac{1}{\lambda_{m_1}^<} \right) 
  \prod_{m_2 = 1}^{|t_1| - 1} \left( \lambda_+ - \lambda_{m_2}^> \right)
  = R e^{{\rm i} \Phi}, \label{eq:app:relation_01}
\end{equation}
where $R$ is a finite real number.

For the limit of $|\varepsilon / \gamma| \ll 1$, the equations of
motion for A- and B-sublattice are decoupled [see
  Eqs. (\ref{eq:EOM_A}) and (\ref{eq:EOM_B})].
For this case, the wavefunctions of A- and B-sublattices under a
boundary are determined separately, in general.
We will show that, for the orthogonal boundary, the wavefunctions of
A- and B-sublattices share a common coefficient for the traveling
modes.
The wavefunctions near the left end are written as the linear
combination of the relevant modes as follows:
\begin{align}
  & \phi_{{\rm A} \ell} 
  = \sum_{\tau' = \pm} c_{{\rm A} \tau'} e^{- \tau' {\rm i} \frac{\Phi}{2}} \lambda_{\tau'}^\ell
  + \sum_{m_1=1}^{|t_2|-1} c_{{\rm A} m_1} \lambda_{m_1}^{< \ell}, \\
  & \phi_{{\rm B} \ell} 
  = \sum_{\tau' = \pm} c_{{\rm B} \tau'} e^{\tau' {\rm i} \frac{\Phi}{2}} \lambda_{\tau'}^\ell
  + \sum_{m_2=1}^{t_1-1} c_{{\rm B} m_2} \frac{1}{\lambda_{m_2}^{> \ell}}.
\end{align}
Note that there are two traveling modes for each sublattice in the
limit of $|\varepsilon / \gamma| \ll 1$.
The factors $e^{\mp \tau' {\rm i} \frac{\Phi}{2}}$ are attached for
the convenience of the discussion.
Here we used that each evanescent mode of B-sublattice at the left end
has a pair with that of A-sublattice at the right end, that is, for
each pair of roots, the relation $\lambda_{\rm A} = 1/\lambda_{\rm B}$
holds where $\lambda_{\rm A}$ is a root of Eq. (\ref{eq:EOM_A_alg})
and $\lambda_{\rm B}$ is the corresponding root of
Eq. (\ref{eq:EOM_B_alg}).
The coefficients are determined by employing boundary conditions.
If $c_{{\rm A} \tau'} = c_{{\rm B} \tau'}$ is satisfied for both
$\tau' = \pm 1$, one of the following relations should be hold in
Eqs. (\ref{eq:standingWave_A}) and (\ref{eq:standingWave_B}):
$c_{+1}=0$ and $\theta_{-1}$ is determined, or, $c_{-1}=0$ and
$\theta_{+1}$ is determined.

Let us consider the orthogonal boundary expressed in
Eqs. (\ref{eq:BC_perp_A}) and (\ref{eq:BC_perp_B}).
If $c_{{\rm A} \tau'} = c_{{\rm B} \tau'}$ is satisfied for both
$\tau' = \pm 1$ for this boundary, the following determinant should be
zero.
\begin{widetext}
  \begin{equation}
    D = 
    \left| 
    \begin{array}{cccccccc}
      e^{- {\rm i} \frac{\Phi}{2}} & e^{{\rm i} \frac{\Phi}{2}} & 1 & \cdots & 1 & 0 & \cdots & 0 \\
      e^{- {\rm i} \frac{\Phi}{2}} \lambda_+^{-1} & e^{{\rm i} \frac{\Phi}{2}} \lambda_-^{-1} & \left( \lambda_1^{<} \right)^{-1} & \cdots & \left( \lambda_{|t_2| - 1}^{<} \right)^{-1}  & 0 & \cdots & 0 \\
      \vdots & \vdots & \vdots & \ddots & \vdots & \vdots & \ddots & \vdots \\
      e^{- {\rm i} \frac{\Phi}{2}} \lambda_+^{-|t_2|+1} & e^{{\rm i} \frac{\Phi}{2}} \lambda_-^{-|t_2|+1} & \left( \lambda_1^{<} \right)^{-|t_2|+1} & \cdots & \left( \lambda_{|t_2| - 1}^{<} \right)^{-|t_2|+1}  & 0 & \cdots & 0 \\
      e^{{\rm i} \frac{\Phi}{2}} & e^{- {\rm i} \frac{\Phi}{2}} & 0 & \cdots & 0 & 1 & \cdots & 1 \\
      e^{{\rm i} \frac{\Phi}{2}} \lambda_+^{-1} & e^{- {\rm i} \frac{\Phi}{2}} \lambda_-^{-1} & 0 & \cdots & 0  & \left( \frac{1}{\lambda_1^{>}} \right)^{-1} & \cdots & \left( \frac{1}{\lambda_{t_1 - 1}^{>}} \right)^{-1} \\
      \vdots & \vdots & \vdots & \ddots & \vdots & \vdots & \ddots & \vdots \\
      e^{{\rm i} \frac{\Phi}{2}} \lambda_+^{-t_1+1} & e^{- {\rm i} \frac{\Phi}{2}} \lambda_-^{-t_1+1} & 0 & \cdots & 0 & \left( \frac{1}{\lambda_1^{>}} \right)^{-t_1+1} & \cdots & \left( \frac{1}{\lambda_{t_1 - 1}^{>}} \right)^{-t_1+1} \\
    \end{array}
    \right|. \label{eq:app:D}
  \end{equation}
  By cofactor expansion and use the relation on the Vandermonde
  matrix, one gets the following relation,
  \begin{align}
    & D 
    = 
    \left( -1 \right)^{ \frac{t_1 (t_1 - 1)}{2} + \frac{|t_2| (|t_2| - 1)}{2} }
    \prod_{1 \le m_1 < m_2 \le |t_2| - 1} \left( \frac{1}{\lambda_{m_1}^<} - \frac{1}{\lambda_{m_2}^<} \right)
    \prod_{1 \le m_1 < m_2 \le t_1 - 1} \left( \lambda_{m_1}^> - \lambda_{m_2}^> \right) \nonumber \\
    & \times 
    \left\{
    e^{{\rm i} \Phi} 
    \prod_{m_1 = 1}^{t_2 - 1} \left( \frac{1}{\lambda_-} - \frac{1}{\lambda_{m_1}^<} \right) 
    \prod_{m_2 = 1}^{|t_1| - 1} \left( \lambda_- - \lambda_{m_2}^> \right)
    -
    e^{-{\rm i} \Phi} 
    \prod_{m_1 = 1}^{t_2 - 1} \left( \frac{1}{\lambda_+} - \frac{1}{\lambda_{m_1}^<} \right) 
    \prod_{m_2 = 1}^{|t_1| - 1} \left( \lambda_+ - \lambda_{m_2}^> \right)
    \right\}. \label{eq:app:D_02}
  \end{align}
\end{widetext}
By using Eq. (\ref{eq:app:relation_01}) in the second line of
Eq. (\ref{eq:app:D_02}), one gets $D=0$.
Therefore, one of the the following two relations should be held for
the orthogonal boundary; $c_{+1}=0$ and $\theta_{-1}$ is determined,
or, $c_{-1}=0$ and $\theta_{+1}$ is determined.


%

\end{document}